# Title: A nearby transiting rocky exoplanet that is suitable for atmospheric investigation

**Authors:** T. Trifonov[1*], J. A. Caballero[2], J. C. Morales[3,4], A. Seifahrt[5], I. Ribas[3,4], A. Reiners[6], J. L. Bean[5], R. Luque[7,8], H. Parviainen[7,8], E. Pallé[7,8], S. Stock[9], M. Zechmeister[6], P. J. Amado[10], G. Anglada-Escudé[3,4], M. Azzaro[11], T. Barclay[12,13], V. J. S. Béjar[7,8], P. Bluhm[9], N. Casasayas-Barris[7,8], C. Cifuentes[2], K. A. Collins[14], K. I. Collins[15], M. Cortés-Contreras[2], J. de Leon[16], S. Dreizler[6], C. D. Dressing[17], E. Esparza-Borges[7,8], N. Espinoza[18], M. Fausnaugh[19], A. Fukui[20,7], A. P. Hatzes[21], C. Hellier[22], Th. Henning[1], C. E. Henze[23], E. Herrero[3,4], S. V. Jeffers[6,24], J. M. Jenkins[23], E. L. N. Jensen[25], A. Kaminski[9], D. Kasper[5], D. Kossakowski[1], M. Kürster[1], M. Lafarga[3,4], D. W. Latham[14], A. W. Mann[26], K. Molaverdikhani[9], D. Montes[27], B. T. Montet[28], F. Murgas[7,8], N. Narita[29,30,31,7], M. Oshagh[7,8], V. M. Passegger[32,33], D. Pollacco[34], S. N. Quinn[14], A. Quirrenbach[9], G. R. Ricker[19], C. Rodríguez López[10], J. Sanz-Forcada[2], R. P. Schwarz[35], A. Schweitzer[32], S. Seager[19,36,37], A. Shporer[19], M. Stangret[7,8], J. Stürmer[9], T. G. Tan[38], P. Tenenbaum[19], J. D. Twicken[39,23], R. Vanderspek[19], J. N. Winn[40]

**Affiliations:**

[1] Max-Planck-Institut für Astronomie, D-69117 Heidelberg, Germany.

[2] Centro de Astrobiología (Consejo Superior de Investigaciones Científicas – Instituto Nacional de Técnica Aeroespacial), E-28692 Villanueva de la Cañada, Madrid, Spain.

[3] Institut de Ciències de l'Espai (Consejo Superior de Investigaciones Científicas), E-08193 Bellaterra, Barcelona, Spain.

[4] Institut d'Estudis Espacials de Catalunya, E-08034 Barcelona, Spain.

[5] Department of Astronomy and Astrophysics, University of Chicago, Chicago, IL 60637, USA.

[6] Institut für Astrophysik, Georg-August-Universität, D-37077 Göttingen, Germany.

[7] Instituto de Astrofísica de Canarias, E-38205 La Laguna, Tenerife, Spain.

[8] Departamento de Astrofísica, Universidad de La Laguna, E-38206 La Laguna, Tenerife, Spain.

[9] Landessternwarte, Zentrum für Astronomie der Universität Heidelberg, D-69117 Heidelberg, Germany.

[10] Instituto de Astrofísica de Andalucía (Consejo Superior de Investigaciones Científicas), E-18008 Granada, Spain.

[11] Centro Astronómico Hispano-Alemán, Observatorio de Calar Alto, E-04550 Gérgal, Almería, Spain.

[12] NASA Goddard Space Flight Center, Greenbelt, MD 20771, USA.

[13] University of Maryland, Baltimore County, Baltimore, MD 21250, USA.

[14] Center for Astrophysics, Harvard & Smithsonian, Cambridge, MA 02138, USA.

[15] Department of Physics and Astronomy, George Mason University, Fairfax, VA 22030, USA.

[16] Department of Astronomy, Graduate School of Science, University of Tokyo, Tokyo 113-0033, Japan.

[17] Astronomy Department, University of California at Berkeley, Berkeley, CA 94720, USA.

[18] Space Telescope Science Institute, Baltimore, MD 21218, USA.

[19] Department of Physics and Kavli Institute for Astrophysics and Space Research, Massachusetts Institute of Technology, Cambridge, MA 02139, USA.

[20] Department of Earth and Planetary Science, Graduate School of Science, University of Tokyo, Tokyo 113-0033, Japan.

[21] Thüringer Landessternwarte Tautenburg, D-07778 Tautenburg, Germany.

[22] Astrophysics Group, Keele University, Staffordshire ST5 5BG, UK.

[23] NASA Ames Research Center, Moffett Field, CA 94035, USA.

[24] Max-Planck-Institut für Sonnensystemforschung, D-37077, Göttingen, Germany.

[25] Department of Physics and Astronomy, Swarthmore College, Swarthmore, PA 19081, USA.

[26] Department of Physics and Astronomy, University of North Carolina at Chapel Hill, Chapel Hill, NC 27599, USA.

[27] Departamento de Física de la Tierra y Astrofísica and Instituto de Física de Partículas y del Cosmos, Facultad de Ciencias Físicas, Universidad Complutense de Madrid, E-28040 Madrid, Spain.

[28] School of Physics, University of New South Wales, Sydney NSW 2052, Australia.

[29] Komaba Institute for Science, University of Tokyo, Tokyo 153-8902, Japan.

[30] Japan Science and Technology Agency, Precursory Research for Embryonic Science and Technology, Tokyo 153-8902, Japan.

[31] Astrobiology Center, Tokyo 181-8588, Japan.

[32] Hamburger Sternwarte, Universität Hamburg, D-21029 Hamburg, Germany.

[33] Homer L. Dodge Department of Physics and Astronomy, University of Oklahoma, Norman, OK 73019, USA.

[34] Department of Physics, University of Warwick, Coventry CV4 7AL, UK.

[35] Patashnick Voorheesville Observatory, Voorheesville, NY 12186, USA.

[36] Department of Earth, Atmospheric and Planetary Sciences, Massachusetts Institute of Technology, Cambridge, MA 02139, USA.

[37] Department of Aeronautics and Astronautics, Massachusetts Institute of Technology, Cambridge, MA 02139, USA.

[38] Perth Exoplanet Survey Telescope, Perth WA 6010, Australia.

[39] Search for Extraterrestrial Intelligence Institute, Mountain View, CA 94043, USA.

[40] Department of Astrophysical Sciences, Princeton University, Princeton, NJ 08544, USA.

*Correspondence to: trifonov@mpia.de

**Abstract:**

Spectroscopy of transiting exoplanets can be used to investigate their atmospheric properties and habitability. Combining radial velocity (RV) and transit data provides additional information on exoplanet physical properties. We detect a transiting rocky planet with an orbital period of 1.467 days around the nearby red dwarf star Gliese 486. The planet Gliese 486 b is 2.81 Earth masses and 1.31 Earth radii, with uncertainties of 5%, as determined from RV data and photometric light curves. The host star is at a distance of ~8.1 parsecs, has a *J*-band magnitude of ~7.2, and is observable from both hemispheres of Earth. On the basis of these properties and the planet's short orbital period and high equilibrium temperature, we show that this terrestrial planet is suitable for emission and transit spectroscopy.

**Main Text**

The combination of transit photometry and Doppler radial velocity (RV) measurements can determine precise values of the masses, radii, bulk densities, and surface gravities of exoplanets. Determination of the exoplanet's atmospheric properties is possible using transmission and emission spectroscopy, but doing so for rocky exoplanets is challenging because of their small size. The Calar Alto high-Resolution search for M dwarfs with Exoearths with Near-infrared and optical Echelle spectrographs (CARMENES) survey (*1*) and combination with the Transiting Exoplanet Survey Satellite (TESS) mission (*2*) together have the sensitivity required to detect and, potentially, jointly investigate and characterise nearby exoplanet systems. Small exoplanets are easier to detect around red dwarfs (main sequence stars of spectral type M), as those stars are themselves small and of low mass. Particularly important are small, Earth-sized terrestrial planets in the habitable zone (*3, 4*), the region where liquid water could exist on the surface. The orbital periods expected for planets in the habitable zone around M dwarfs are a few tens of days, and the predicted RV signals are large enough to be detectable.

M dwarfs are abundant in the Solar neighborhood; of the 357 cataloged main-sequence stars within 10 pc of the Sun, 283 (79%) are of type M (5, 6). Nearby exoplanets are favored for follow-up characterization, mainly because of their brighter host stars (producing a higher signal-to-noise ratio). Within 10 pc, ~80 planets in 40 stellar systems are known, of which ~50 planets orbit around 35 M dwarf hosts. These include the closest exoplanet systems, such as Proxima Centauri (7, 8) and Barnard's star (9).

We observed the nearby star Gliese 486 [Wolf 437, TESS Object of Interest (TOI) 1827], a red dwarf of spectral type M3.5 V, as one of the ~350 targets in the CARMENES survey (10). RV monitoring of the star between 2016 and early 2020 showed a periodicity of 1.467 days with a false-alarm probability of <0.1% (11). No counterpart was found in stellar activity indices, suggesting that the signal was due to an orbiting planet rather than stellar variability, which is common in M dwarfs. We used photometric data from TESS to confirm the presence of the planet, identifying 13 transit events with a periodicity of 1.467 days (11). At a distance of 8.1 pc, Gliese 486 is the third-closest transiting exoplanet system known, and Gliese 486 b is the closest transiting planet around a red dwarf with a measured mass.

We list the physical properties of the star Gliese 486 and planet Gliese 486 b in Table 1 (11). From the CARMENES spectroscopic observations and a photometric data compilation (12), we computed a stellar radius of $0.328 \pm 0.011$ solar radii ($R_\odot$) and a mass of $0.323 \pm 0.015$ solar masses ($M_\odot$) following (13). Because of its closeness, Gliese 486 has been a target of direct-imaging exoplanet searches (14, 15), which placed upper limits on low-mass stellar and substellar companions at sky-projected physical separations between 1.2 and 161 astronomical units (au), larger than the orbit we find for Gliese 486 b.

We supplemented the TESS photometry with ground-based photometric monitoring and archival time series data to further characterize the transit events and determine the stellar rotation period. Using photometry of Gliese 486 collected by the Wide Angle Search for Planets (WASP) (16) between 2008 and 2014 and by the All-Sky Automated Survey for Supernovae (ASAS-SN) (17) between 2012 and 2020, we measured a stellar rotation period $P_{rot} = 130.1^{+1.6}_{-1.2}$ days, which is consistent with our expectations for an old and weakly active M-dwarf star and much longer than the planet orbital period (fig. S4). We observed two additional transit events using the Multicolor Simultaneous Camera for studying Atmospheres of Transiting exoplanets 2 (MuSCAT2) (18) at the 1.5-m Telescopio Carlos Sánchez at Observatorio del Teide on 9 May 2020 and 12 May 2020 and three more transits with the 1.0-m Las Cumbres Observatory Global Telescope (LCOGT) (19) at Siding Spring Observatory on 15 May 2020, 24 May 2020, and 5 June 2020.

We complemented our CARMENES RV observations of Gliese 486 with data from the M-dwarf Advanced Radial velocity Observer Of Neighboring eXoplanets (MAROON-X) spectrograph (20) at the 8.1-m Gemini North telescope. In total, we obtained 80 CARMENES spectra between 2016 and 2020 and 65 with MAROON-X between May and June 2020. These data provide complete phase coverage of the Gliese 486 b RV signal (Fig. 1), with a total weighted root mean square residual of 1.05 m s$^{-1}$.

We performed an orbital analysis using the EXO-STRIKER oftware (21). Global parameter optimization was performed by simultaneously fitting Keplerian orbit models to the CARMENES visual channel (VIS), MAROON-X red and blue channels, and the TESS photometry. An alternative model that also includes transit data from MuSCAT2 and LCOGT provides consistent results (11). For Gliese 486 b, we obtained a planet orbital period $P_b = 1.467119^{+0.000031}_{-0.000030}$ days and orbital inclination $i_b = 88.4^{+1.1}_{-1.4}$ degrees. Using the RV semiamplitude $K_b = 3.37^{+0.08}_{-0.08}$ m s$^{-1}$, the stellar parameters of Gliese 486, and the orbital parameters, we derived a dynamical planet mass $M_b = 2.82^{+0.11}_{-0.12}$ Earth masses (M$_E$), a semimajor axis $a_b = 0.01732^{+0.00027}_{-0.00027}$ au, and a planet radius $R_b = 1.306^{+0.063}_{-0.067}$ Earth radii (R$_E$). We concluded that Gliese 486 b has a circular orbit with an upper limit on the eccentricity $e_b < 0.05$ at 68.3% confidence level. This low eccentricity is consistent with the short orbital period, as star-planet tidal forces would act to circularize the orbit. We performed star-planet tidal simulations of the Gliese 486 system with the EQTIDE integrator (22) and found that the orbit of Gliese 486 b becomes fully circularized within ~1 million years.

From the planet mass and radius, we derived a planet bulk density $\rho_b = 7.0^{+1.2}_{-1.0}$ 10$^3$ kg m$^{-3}$ (~1.3 times that of Earth) and a surface gravity $g_b = 16.2^{+1.9}_{-1.6}$ m s$^{-2}$ (~1.7 times that of Earth), respectively. From the location of Gliese 486 b in a radius-mass diagram (Fig. 2), its density indicates an iron-to-silicate ratio similar to Earth's (23). The inferred mass and radius of about 2.82 M$_E$ and 1.31 R$_E$ put Gliese 486 b at the boundary between Earth and super-Earth planets

(24), but the bulk density indicates a massive terrestrial planet rather than an ocean planet (25). The escape velocity at 1 $R_b$ is $v_e = 16.4^{+0.6}_{-0.5}$ km s$^{-1}$. For an energy-limited atmospheric atmospheric escape model (26) and the previously measured host star x-ray flux upper limit (27), we derive a low photoevaporation rate of $M_{phot} < 10^7$ kg s$^{-1}$. From the stellar bolometric luminosity and the planet semimajor axis, we inferred a planet irradiance $S_b = 40.3^{+1.5}_{-1.4}$ times that of Earth. Assuming complete absorbance (a Bond albedo $A_B = 0$), this equates to an equilibrium temperature $T_{eq} = 701^{+13}_{-13}$ K, which is slightly cooler than that of Venus.

Fig. 3 shows how Gliese 486 b compares with other possibly rocky planets around nearby M dwarfs (those with measured masses and radii $R_p < 2.0$ $R_E$) using standard metrics for transmission and emission spectroscopy. Figure 3A shows the expected primary transit transmission signal d per atmospheric scale height $H$ ($\delta \approx 2 H R_p / R_\star^2$, where $R_p$ is the radius of the planet, and $R_\star$ is the radius of the star) as a function of apparent magnitude in the $K_s$ band.

Figure 3B shows the transmission spectroscopy metric as a function of $T_{eq}$, whereas panel Fig. 3C shows the emission spectroscopy metric, which is the signal-to-noise ratio expected for a single secondary eclipse observation by the James Webb Space Telescope (28). Figure 3, B and C, show planets around M dwarfs with measured masses. With a radius of 1.31 $R_E$, Gliese 486 b is located well below the radius range of 1.4 to 1.8 $R_E$, under which planets are expected to have lost their primordial hydrogen-helium atmospheres owing to photoevaporation processes (29). It remains unknown how stellar irradiation and planet surface gravity affect the formation and retention of secondary atmospheres. Planets with $T_{eq} > 880$ K, such as 55 Cancri e (30), are expected to have molten (lava) surfaces and no atmospheres, except for vaporized rock (31). Gliese 486 b is not hot enough to be a lava world, but its temperature of ~700 K makes it suitable for emission spectroscopy and phase curve studies in search of an atmosphere (28). Our orbital model constrains the secondary eclipse time to within 13 min (at 1$\sigma$ uncertainty), which is necessary for efficient scheduling of observations. Compared with other known nearby rocky planets around M dwarfs, Gliese 486 b has a shorter orbital period and correspondingly higher equilibrium temperature of ~700 K and orbits a brighter, cooler, and less active stellar host.

**Acknowledgments:** Based on observations made with the CARMENES spectrograph at the 3.5 m telescope of the Centro Astronómico Hispano-Alemán de Calar Alto (CAHA, Almería, Spain), funded by the German Max-Planck-Gesellschaft (MPG), the Spanish Consejo Superior de Investigaciones Científicas (CSIC), the European Regional Development Fund, and the CARMENES Consortium members, and stored at the CARMENES data archive at CAB (INTA-CSIC); the MAROON-X spectrograph, which was funded by the David and Lucile Packard Foundation, the Heising-Simons Foundation, the Gemini Observatory, and the University of Chicago (the MAROON-X team thanks the staff of the Gemini Observatory for their assistance with the commissioning and operation of the instrument); the LCOGT network; the MuSCAT2 instrument, developed by the Astrobiology Center, at Telescopio Carlos Sánchez operated on the island of Tenerife by the Instituto de Astrofísica de Canarias in the Spanish Observatorio del Teide; and data collected by the TESS mission.

**Funding:** Deutsche Forschungsgemeinschaft through research unit FOR2544 "Blue Planets around Red Stars" and priority program SPP1992 "Exploring the Diversity of Extrasolar Planets", Agencia Estatal de Investigación of the Ministerio de Ciencia e Innovación and the European Regional Development Fund through projects PID2019-109522GB-C51/2/3/4, PGC2018-098153-B-C33, SEV-2017-0709, MDM-2017-0737, AYA2016-79425-C3-1/2/3-P, ESP2016-80435-C2-1-R, SEV-2015-0548, Klaus Tschira Stiftung, European Union's Horizon 2020 through Marie Sklodowska Curie grant 713673, "la Caixa" through INPhINT grant LCF/BQ/IN17/1162033, NASA through grants NNX17AG24G, 80NSSC19K0533, 80NSSC19K1721, 80NSSC18K158 and the NASA Science Mission Directorate, Japan Society for the Promotion of Science KAKENHI through grants JP17H04574, JP18H01265 and JP18H05439, Japan Science and Technology Agency PRESTO through grant JPMJPR1775.

**Author contributions**:

T.T. analyzed and interpreted the data and wrote the manuscript. J.A.C. is the CARMENES instrument astronomer and helped write the manuscript. J.C.M. scheduled the CARMENES observations, and performed preliminary CARMENES RV analysis and transit predictions. I.R. and A.R. are the CARMENES project scientists and identified the periodic radial velocity signal. J.L.B., R.L., E.P., S.St, and K.M. contributed to writing the manuscript. A.Se. and M.Z. reduced the CARMENES and MAROON-X spectra. G.A.E., P.B., N.E., A.P.H analyzed the RV and photometric time series. C.C., M.C.C., V.M.P., J.S.F. and A.Sc. determined stellar parameters. The following authors contributed to instrument operations, science coordination and data analysis: P.J.A., M.A., V.J.S.B., S.D., T.H., S.V.J., A.K., D.Ko., M.K., M.L., D.M., A.Q., C.R.L. for CARMENES, D.Ka., B.T.M., J.S. for MAROON-X, N.C.B., J.d.L., E.E.B., A.F., F.M., N.N., H.P., M.S. for MuSCAT2, K.A.C., K.I.C., E.L.N.J., A.Sh., R.P.S for LCOGT, C.H., E.H., D.P., T.G.T. for PEST, TJO, and SuperWASP**,** T.B., C.D.D. M.F., C.E.H., J.M.J., D.W.L., A.W.M., S.N.Q., G.R.R., S.Se., P.T., J.D.T., R.V., J.N.W. for TESS.

**Competing interests:**

We declare no competing interests.

**Data and materials availability:** The Exo-Striker code is freely available at https://github.com/3fon3fonov/exostriker. The juliet code is freely available at https://github.com/nespinoza/juliet. The EqTide code is freely available at https://github.com/RoryBarnes/EqTide. All time series used in this work (RV, activity indices and light curves from CARMENES, MAROON-X, HARPS, HIRES, MuSCAT2, LCOGT, SuperWASP, ASAS-SN, TJO, and TESS) are available at the Centro de Astrobiología CARMENES data archive http://carmenes.cab.inta-csic.es/ as machine-readable tables in GJ486.zip.

**Supplementary Materials:**

Materials and Methods
Figures S1-S10
Tables S1-S6
References (32-102)
Data S1

**Table 1. Measured properties of Gliese 486 and its planet.** We used $G = 6.67430 \; 10^{-11}$ m$^3$ kg$^{-1}$ s$^{-2}$, $M_{sol} = 1.98847 \; 10^{30}$ kg, $R_{sol} = 6.957 \; 10^8$ m, $M_E = 5.9722 \; 10^{24}$ kg, $R_E = 6.3781 \; 10^6$ m. The tabulated rotation period is a proxy obtained from a quasi-periodic representation of the photometric variability. The eccentricity upper limit of <0.05 is constrained at the 68.3% confidence level. The tabulated equilibrium temperature would be 60 K cooler if the Bond albedo were 0.30.

| Stellar parameters | Value |
| --- | --- |
| Right ascension (J2000 equinox) | 12:47:56.62 |
| Declination (J2000 equinox) | +09:45:05.0 |
| Spectral type | M3.5±0.5 V |
| $J$-band magnitude (mag) | 7.195±0.026 |
| Mass ($M_{sol}$) | 0.323±0.015 |
| Radius ($R_{sol}$) | 0.328±0.011 |
| Luminosity ($L_{sol}$) | 0.01210±0.00023 |
| Effective temperature (K) | 3340±54 |
| Distance (pc) | 8.0761±0.0041 |
| Rotation period (d) | $130.1^{+1.6}_{-1.2}$ |
| Metallicity [Fe/H] (dex) | +0.07±0.16 |
| **Planetary parameters** | **Value** |
| Orbital period (d) | $1.467119^{+0.000031}_{-0.000030}$ |
| Radial velocity semi-amplitude (m s$^{-1}$) | $3.370^{+0.078}_{-0.080}$ |
| Eccentricity | <0.05 |
| Argument of periastron (deg) | undefined |
| Time of inferior transit (barycentric Julian date, BJD) | $2458931.15935^{+0.00042}_{-0.00042}$ |
| Orbital semi-major axis (au) | $0.01734^{+0.00026}_{-0.00027}$ |
| Mass ($M_E$) | $2.82^{+0.11}_{-0.12}$ |
| Radius ($R_E$) | $1.305^{+0.063}_{-0.067}$ |
| Inclination (deg) | $88.4^{+1.1}_{-1.4}$ |
| Insolation ($S_E$) | $40.3^{+1.5}_{-1.4}$ |
| Mean density ($10^3$ kg m$^{-3}$) | $7.0^{+1.2}_{-1.0}$ |
| Surface gravitational acceleration (m s$^{-2}$) | $16.4^{+0.6}_{-0.5}$ |
| Equilibrium temperature (K) | $701^{+13}_{-13}$ |

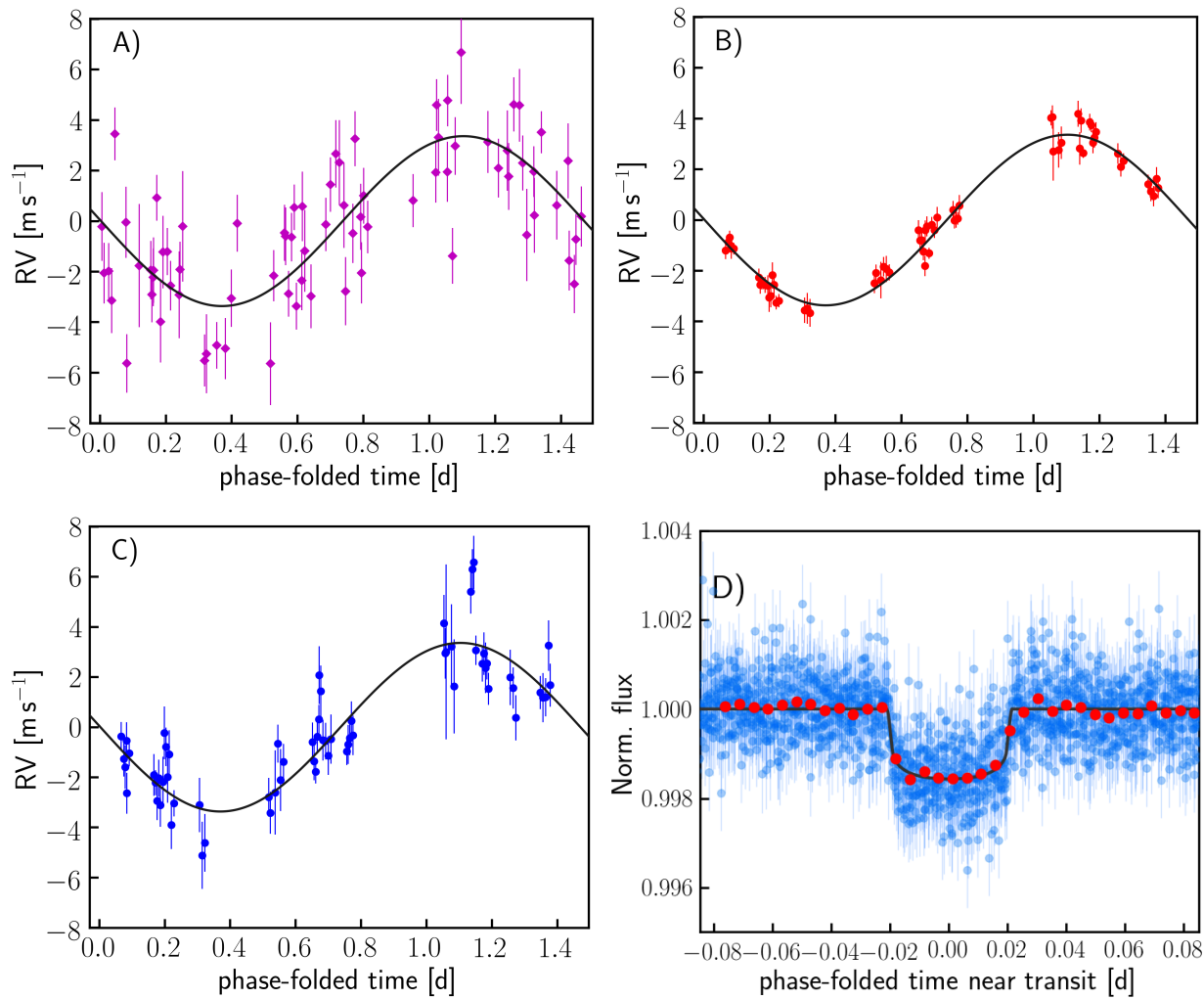

**Fig. 1. Radial velocity and light curves of Gliese 486.** Phase-folded RV data from (**A**) CARMENES VIS, (**B**) MAROON-X red, and (**C**) MAROON-X blue, and (**D**) TESS photometric data. Blue circles in (**D**) represent the phase-folded 2-min cadence TESS transit photometry, whereas red circles are 1-hour bins of the phase-folded data. Error bars indicate 1s uncertainties of individual measurements. Black solid curves in all panels are the maximum likelihood orbital model from a joint fitting of all these data simultaneously. Norm. flux, normalized flux; d, days.

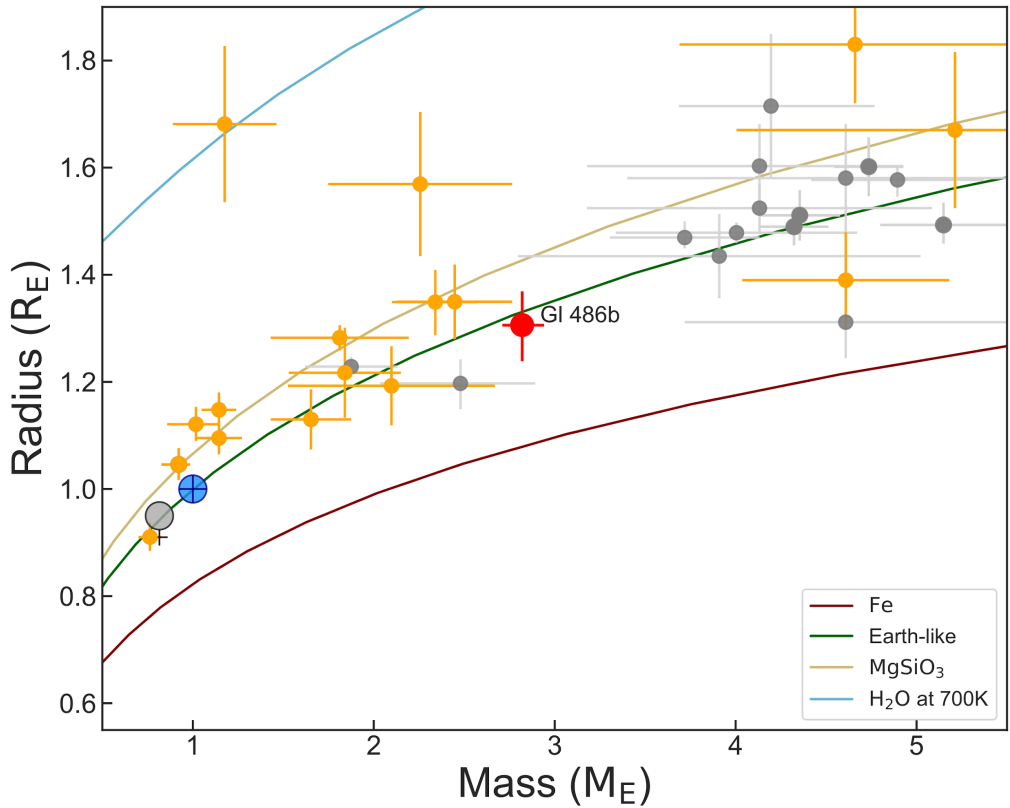

**Fig. 2. Mass-radius diagram for known transiting planets with measured masses between 0.5 $M_E$ and 5.5 $M_E$ and radii between 0.5 $R_E$ and 2.0 $R_E$.** We show all cases with precision better than 30% (see supplementary text). Gliese 486 b is shown in red, planets orbiting around late-type stars with $T_{eff}$ < 4000 K are shown in orange, and hotter stars are shown in dark gray. Earth (blue circle with cross) and Venus (light gray circle with ♀ symbol) are shown for comparison. Curves show theoretical planet mass-radius relationships for compositions indicated in the legend: pure water ($H_2O$), pure enstatite ($MgSiO_3$) rock, an Earth-like mixture of 50% enstatite and 50% iron, and pure iron (Fe) (*23*).

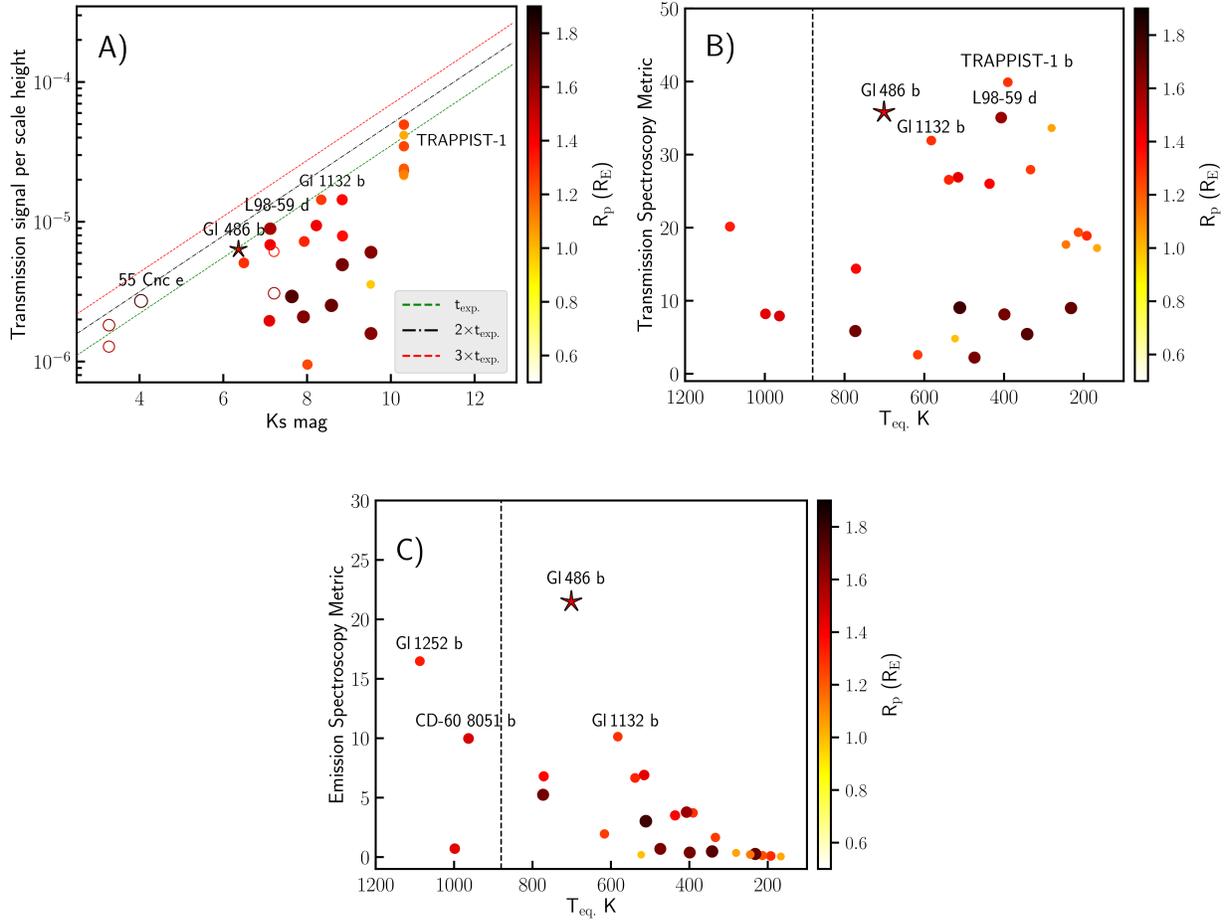

**Fig. 3. Metrics for transmission and emission spectroscopy for rocky planets with measured mass around nearby M dwarfs. (A)** Expected primary transit transmission signal per scale height as a function of $K$s-band magnitude. Gliese 486 b is shown with a star, planets around bright G and K dwarfs at a distance of < 30 pc are shown with open circles, and planets around M dwarfs are shown with solid circles. The color bar indicates the planet radius. Selected planets are labeled. **(B)** Same as **(A)**, but for the transmission spectroscopy metric (computed homogeneously with a scale factor 0.190) as a function of $T_{eq}$. **(C)** Same as **(A)**, but for the emission spectroscopy metric as a function of $T_{eq}$. In **(A)**, diagonal dashed lines mark expected amplitudes of spectral features in transmission at three different, arbitrary exposure times $t_{exp}$, $2t_{exp}$, and $3t_{exp}$ with the same instrumental setup. In **(B)** and **(C)**, planets hotter than the vertical

lines at $T_{eq}$ = 880 K are expected to have molten lava surfaces. See supplementary text for details.

# Science
## AAAS

Supplementary Materials for

# A nearby transiting rocky exoplanet that is suitable for atmospheric investigation

T. Trifonov, J. A. Caballero, J. C. Morales, A. Seifahrt, I. Ribas, A. Reiners, J. L. Bean, R. Luque, H. Parviainen, E. Pallé, S. Stock, M. Zechmeister, P. J. Amado, G. Anglada-Escudé, M. Azzaro, T. Barclay, V. J. S. Béjar, P. Bluhm, N. Casasayas-Barris, C. Cifuentes, K. A. Collins, K. I. Collins, M. Cortés-Contreras, J. de Leon, S. Dreizler, C. D. Dressing, E. Esparza-Borges, N. Espinoza, M. Fausnaugh, A. Fukui, A. P. Hatzes, C. Hellier, Th. Henning, C. E. Henze, E. Herrero, S. V. Jeffers, J. M. Jenkins, E. L. N. Jensen, A. Kaminski, D. Kasper, D. Kossakowski, M. Kürster, M. Lafarga, D. W. Latham, A. W. Mann, K. Molaverdikhani, D. Montes, B. T. Montet, F. Murgas, N. Narita, M. Oshagh, V. M. Passegger, D. Pollacco, S. N. Quinn, A. Quirrenbach, G. R. Ricker, C. Rodríguez López, J. Sanz-Forcada, R. P. Schwarz, A. Schweitzer, S. Seager, A. Shporer, M. Stangret, J. Stürmer, T. G. Tan, P. Tenenbaum, J. D. Twicken, R. Vanderspek, J. N. Winn

Correspondence to: trifonov@mpia.de

**This PDF file includes:**

Materials and Methods
Figs. S1 to S10
Tables S1 to S6

## Materials and Methods

### Spectroscopic observations

Spectroscopic data employed for the RV analysis were obtained with the CARMENES spectrograph, the newly commissioned MAROON-X spectrograph, and from archival data from the High Resolution Echelle Spectrometer (HIRES) (32) at the 10.0 m Keck I Telescope, and the High Accuracy Radial velocity Planet Searcher (HARPS) (33) at the European Southern Observatory (ESO) 3.6 m Telescope. Fig. S1 shows the available RV data combined after subtraction of the mean RV offset.

Gliese 486 (34) is one of the about 350 M-dwarf targets regularly monitored in the CARMENES (Calar Alto high-Resolution search for M dwarfs with Exo-earths with Near-infrared and optical Echelle Spectrographs) guaranteed time observation program. Detailed descriptions of the CARMENES instrument at the 3.5 m Calar Alto telescope and the on-going exoplanet survey can be found in (*1*) and (10). For Gliese 486 we obtained 80 pairs of optical (VIS: 520-960 nm) and near-infrared (NIR: 960-1710 nm) spectra between January 2016 and June 2020 with a total time baseline of 1612.7 d. The typical exposure time was about 20 min, chosen with the goal of reaching a signal-to-noise ratio (S/N) of 150 in the *J* band. All the spectra went through the standard CARMENES data flow (3*5*). Using the version 2.20 of the data reduction pipeline and of the SpEctrum Radial Velocity AnaLyser (SERVAL) (36), we computed VIS and NIR radial velocity (RV) measurements. Additionally, we computed and corrected the nightly zero-point (NZP) offsets of the CARMENES data (3*7*). Four CARMENES epochs were discarded because the spectra were taken without simultaneous Fabry-Pérot etalon wavelength calibration. The resulting 76 VIS RVs had a weighted root-mean-square velocity, wrms$_{C-VIS}$, of 2.56 m s$^{-1}$ and a median uncertainty, $\hat{\sigma}_{C-VIS}$, of 1.17 m s$^{-1}$. We additionally discarded 16 NIR spectra obtained before the start of the nominal operations of the NIR channel (3*8*). For the remaining 60 CARMENES NIR measurements of Gliese 486 we measured wrms$_{C-NIR}$ = 6.36 m s$^{-1}$ and $\hat{\sigma}_{C-NIR}$ = 4.36 m s$^{-1}$. Simultaneously with the RVs extraction from CARMENES spectra, SERVAL computes the time series of several stellar activity indices: the chromatic index (CRX), the differential line width (dLW), calcium infrared triplet (Ca IRT), Hα, and Na I D1 and D2. Using the RACOON pipeline (39), from the CARMENES spectra we also calculated the full-width half-maximum (FWHM) of the cross-correlation function (CCF) profile, the bisector inverse slope (BIS) span, and contrast stellar line measurement (CON) of the spectral lines (40). The time series of the RV and all activity indices from CARMENES VIS and NIR channels, together with their individual uncertainties, are listed in Tables S1 and S2, respectively.

We also performed RV observations of Gliese 486 using the MAROON-X instrument (20, *41*) on the 8.1 m Gemini North telescope. MAROON-X is a fiber-fed double-channel optical (blue: 500-670 nm, red: 650-920 nm) spectrograph with a resolving power R = 85,000 designed for RV observations of M dwarfs. We obtained 65 spectra of Gliese 486 in 17 visits over 13 nights between 20 May and 02 June 2020 using MAROON-X. Visits comprised between two and six consecutive exposures of 300 or 600 s each, depending on seeing conditions and cloud coverage. The typical S/N per pixel was about 120 and 280 in the blue and red channels, respectively. Each spectral resolution element is sampled by 3.2 pixels on average. The MAROON-X data were reduced using a custom Python 3 pipeline based on tools previously developed for the CRyogenic high-resolution InfraRed Echelle Spectrograph (CRIRES) (*42, 43*). The MAROON-X data reduction software, which is being incorporated into Gemini's data reduction platform, can meanwhile be provided upon reasonable request. Similarly to CARMENES, the MAROON-X wavelength calibration strategy used stabilized Fabry-Pérot etalon exposures that were taken simultaneously with the data using a dedicated fiber. The instrumental drift correction was part of the wavelength calibration. Radial velocities and activity indices were measured using SERVAL. MAROON-X red data have wrms$_{MX-red}$ = 2.26 m s$^{-1}$ and $\hat{\sigma}_{MX-red}$ = 0.39 m s$^{-1}$, and blue data have wrms$_{MX-blue}$ = 2.36 m s$^{-1}$ and $\hat{\sigma}_{MX-blue}$ = 0.82 m s$^{-1}$. The time series of the RV and all activity indices from MAROON-X red and blue, together with their individual uncertainties, are listed in Tables S3 and S4, respectively.

We retrieved archival RV measurements of Gliese 486 taken with HIRES and HARPS. There are 27 HIRES RVs in a published catalogue (*44*), with later NZP corrections (*45*). For Gliese 486 these datasets (*44*) and (*45*) are almost identical, but we decided to use the corrected data set for consistency with the CARMENES data (*37*). The HIRES observations of Gliese 486 were taken between January 1998 and January 2011, with a total temporal baseline of 4740.8 d. After removing an obvious outlier at barycentric Julian date BJD = 2452006.986 with a 3σ-clipping filter, the HIRES RV data have a wrms$_{HIRES}$ = 6.64 m s$^{-1}$ and a $\hat{\sigma}_{HIRES}$ = 3.22 m s$^{-1}$, which are larger than those of CARMENES and MAROON-X. There are 12 NZP-corrected HARPS RVs of Gliese 486 in the HARPS-RVBank database (*46*). The corresponding spectra were taken between June 2004 and May 2011 with a total temporal baseline of 2533.0 d. The HARPS RV data have wrms$_{HARPS}$ = 3.33 m s$^{-1}$ and $\hat{\sigma}_{HARPS}$ = 1.16 m s$^{-1}$. HARPS-RVBank also tabulates CRX, dLW, Ca IRT, Hα, and Na I D1 and D2 computed with SERVAL and FWHM, BIS, and CON computed with the Data Reduction Software (DRS), the standard HARPS pipeline.

Photometric monitoring

Gliese 486 (TOI-1827) was observed in 2 min short-cadence integrations by the TESS spacecraft in Sector 23, camera 1, detector chip number 3, between 18 March 2020 and 16 April 2020. We retrieved the TESS data from the Mikulski Archive for Space Telescopes. For this target, the Science Processing Operations Center (SPOC) (*47*) provided both simple aperture photometry (SAP) and systematics-corrected photometry adapted from the *Kepler* Pre-search Data Conditioning algorithm (PDC) (*48, 49*). The PDC light curve is constructed by detrending the SAP light curve using a linear combination of cotrending basis vectors, which are derived from a principal component decomposition of the light curves individually for each sector, camera, and CCD. PDC light curves are corrected for contamination from nearby stars and instrumental systematics including pointing drifts, focus changes, and thermal transients. Fig. S2 shows the target pixel file (TPF) image of Gliese 486 constructed from TESS and *Gaia* DR2 data with the TPFPLOTTER tool (*50*), and a false-color image from *u'*-, *i'*-, *z'*-band Sloan Digital Sky Survey (SDSS9) data (*51*) with the Aladin sky atlas (*52*). Comparing Fig. S2 to previous adaptive optics (*14*) and *Hubble Space Telescope* high-resolution imaging, we expect negligible flux dilution by stellar contaminants in the TESS aperture mask in the epoch of TESS observations [and all photometric observations described below (*53*)].

We carried out additional ground-based photometric monitoring and retrieved archival magnitude series for ruling out nearby eclipsing binaries, further characterizing the transit events, and trying to determine the stellar rotation period. Three transits of Gliese 486 b were observed simultaneously in *g*, *r*, *i*, and *zs* bands with the Multicolour Simultaneous Camera for studying Atmospheres of Transiting exoplanets 2 (MuSCAT2) (*18*) on the 1.52 m Telescopio Carlos Sánchez at Observatorio del Teide on 9 May 2020, 12 May 2020, and 3 June 2020. The observations on 9 May covered 1.7 h centered around the expected transit mid-time, with airmass varying from 1.2 at the beginning of the observations to 1.8 at the end of the observations. The observations on 12 May covered 4.3 h approximately centered around the expected transit mid-time with airmass covering values from 1.05 to 1.45. The observations on 3 June were affected by poor weather conditions, so they were not used. All MuSCAT2 observations were defocused, optimizing the photometry for a star as bright as Gliese 486. However, the lack of suitably bright comparison stars in the field of view led to a sub-optimal photometry, and the white noise estimates in the reduced light curves vary from ~ 2.3 ‰ in *g* to ~ 1.6 ‰ in *zs*. We performed relative photometry using standard aperture photometry calibration and reduction steps with a dedicated MuSCAT2 photometry pipeline based on PyTransit (*54, 55*). The pipeline calculates aperture photometry for a set of comparison stars and aperture sizes, and produces the final relative light curves via global optimization of a model that aims to find the optimal comparison stars and their aperture size while simultaneously modeling the transit and baseline variations as linear combinations of a set of covariates.

We observed three full transits of Gliese 486 with Las Cumbres Observatory Global Telescope (LCOGT) 1.0 m network (*19*) in the *z* filter on 15 May 2020, 24 May 2020, and 5 June 2020. The telescopes are equipped with 4k × 4k cameras having an image scale of 0.389 arcsec pixel$^{-1}$, resulting in a 26 × 26 arcmin$^2$ field of view. The telescopes were defocused and yielded point spread functions with FWHM of approximately 8 arcsec. The transits on 15 May 2020 and 24 May 2020 were observed continuously for 235 and 187 min from the LCOGT node at the South African Astronomical Observatory using 25 s exposures, which resulted in 240 and 171 images, respectively. The transit on 5 June 2020 was observed continuously for 247 min from the LCOGT node at Siding Spring Observatory using 25 s exposures, which resulted in 251 images. The images were calibrated by the standard LCOGT banzai pipeline (*56*) and the photometric data were extracted using the AstroImageJ software package (*57*). Circular apertures with radius 25, 30, and 20 pixels were used to extract differential photometry from the 15 May 2020, 24 May 2020, and 5 June 2020 data, resulting in model residuals of 660, 350, 380 ppm in 10 min bins, respectively.

We observed a full transit of Gliese 486 b continuously for 258 min on 08 Jun 2020 in *Rc* band with the Perth Exoplanet Survey Telescope (PEST) near Perth, Australia. The 0.3 m telescope is equipped with a 1.5k × 1k camera with an image scale of 1.2 arcsec pixel$^{-1}$, resulting in a 31 × 31 arcmin$^2$ field of view. The images had typical stellar point spread functions with a FWHM of 4.0 arcsec. The data did not detect the transit, but did rule out nearby eclipsing binaries in all six stars within 2.5 arcmin of the target that are bright enough to contaminate the TESS data.

The Wide Angle Search for Planets (WASP) transit search consisted of two wide-field arrays of eight cameras, with SuperWASP-North being at the Observatorio del Roque de Los Muchachos in La Palma, Spain, and WASP-South being at the South African Astronomical Observatory in Sutherland, South Africa (*16*). The field of Gliese 486 was observed by both arrays. SuperWASP-North observed Gliese 486 in four consecutive seasons from 2008 to 2011, for spans between 50 and 120 d each season. It was equipped with a 200 mm f/1.8 lens with a broadband filter spanning 400-700 nm, backed by 2k × 2k CCDs, giving a plate scale of 13.7 arcsec pixel$^{-1}$. Observations on every clear night rastered available fields with a typical 15 min cadence. In 2013 and 2014, Gliese 486 was observed by WASP-South for spans of 120 and 170

d. The array was then equipped with 85 mm f/1.2 lenses with an SDSS *r'* filter, giving a plate scale of 32 arcsec pixel$^{-1}$. In the magnitude range of Gliese 486, SuperWASP-North, with its bigger lens and finer plate scale, provided less red noise and better background subtraction than WASP-South. In total, we collected over 51 714 SuperWASP photometric measurements of Gliese 486 from the Northern (wrms = 0.012 mag) and Southern (wrms = 0.051 mag) hemispheres. For comparison purposes and monitoring of systematics, we also collected the light curves of four nearby stars with similar brightness. These stars were: 1SWASP J124802.97+094759.9, *V*=12.95 mag., 1SWASP J124816.33+095108.4, *V*=12.58 mag., BD+10 2472, *V*= 9.70 mag., and TYC 882-378-1, *V*=11.34 mag.

We searched for public time series data of wide-area photometric surveys and databases following (*58*). The sparse All-Sky Automated Survey ASAS (*59*) and Northern Sky Variability Survey NSVS (*60*) data sets of Gliese 486 with rms of 0.066 mag and 0.032 mag, respectively, did not have any significant peak with <0.1% FAP in the periodograms. We also retrieved light curves from the All-Sky Automated Survey for SuperNovae (ASAS-SN) (*61*) in the *g'* and *V* bands, which spanned from November 2012 to May 2020. Because Gliese 486 has a high proper motion, we obtained the *V*- and *g'*-band magnitudes from ASAS-SN by season. We retrieved the calculated real-time magnitudes using aperture photometry centered on the expected equatorial coordinates of Gliese 486 at the middle of every observing season (mid March). The ASAS-SN *V*- and *g'*-band magnitudes are zero-point calibrated with the American Association of Variable Star Observers Photometric All Sky Survey APASS catalogue (*62*). In total, we retrieved 2175 archival data points, of which 984 were taken in the *V* band (972 useful, wrms = 0.020 mag) and 1191 in the *g'* band (1064 useful, wrms = 0.039 mag).

We conducted observations with the 0.8 m Telescopi Joan Oró (TJO) at the Observatori Astronòmic del Montsec in Lleida, Spain, as part of the CARMENES photometric follow-up program. We aimed to cover the ±3σ phase window around the conjunction time predicted by the RV solution at the time of observations. The transit time 1σ uncertainty of 2.35 h implied monitoring Gliese 486 over a time window of 7 h at both sides of the predicted zero phase. We collected data on 9, 11, and 14 April, and 3 May 2020, obtaining a total of 1578 images with the Johnson *R* filter using the Large Area Imager for Astronomy (LAIA) imager, a 4k × 4k CCD with a field of view of 30 arcmin and a scale of 0.4 arcsec pixel$^{-1}$. The images were calibrated with dark, bias, and flat fields frames using the observatory pipeline. Differential photometry was extracted with *AstroImageJ* using the aperture size and the set of comparison stars selected to minimize the rms of the photometry. We covered most of the early side of the foreseen time window, including the predicted transit epoch. However, no transit was detected. The TESS data later showed the transit occured 2.04 h later than we had initially predicted (but within the 1σ uncertainty at that time), corresponding to an orbital phase that had not been sampled.

Stellar parameters and rotation period

Stellar parameter estimates for Gliese 486 are given in Table 1. Published spectral types of Gliese 486 have varied between M3.0 V (*63*) and M4.0 V (*64*), i.e., a spectral typing uncertainty of 0.5 subtypes (*65*). The photosphere parameters ($T_{eff}$, log *g*, and [Fe/H]) of Gliese 486 were adopted from previous compilations by (*66*) that used CARMENES spectra. The bolometric luminosity was taken from (*12*) and the $T_{\text{eff}}$ from (*66*); combining these with the Stefan-Boltzmann law, we calculated the stellar radius. The mass-radius relation of (*13*) was used to determine the stellar mass.

Gliese 486 is an M dwarf with very weak chromospheric activity (*67,68,69*). It is a slow rotator with very narrow spectroscopic lines (*70, 71*), faint Ca II H&K emission (*72, 73*), and weak magnetic field (*74*). A log $R'_{\text{HK}}$ was calculated by averaging the HIRES $S_{\text{MWO}}$ index series after discarding three obvious outliers and a fourth datum with a low S/N. The mean $S_{\text{MWO}}$ corresponds to log $R'_{\text{HK}}$ = –5.51±0.39 and an expected rotation period of ~90 d (using the relations of (*70*) and the $V$ and $K$s magnitudes of (*75*) and (*76*), respectively). The mean value of log $R'_{\text{HK}}$ from the HIRES data is higher than that from HARPS data (*70*), but consistent within 1σ, and the larger uncertainty arises from intrinsic variability of the Ca II H&K doublet.

We used the photometric data sets of SuperWASP and ASAS-SN to measure the stellar rotation period of Gliese 486. After accounting for the discrete Fourier transform window functions of the observations, three significant peaks appear in the periodograms (Fig. S3), at approximately 189 d, 125 d, and 93 d, similar to the 1/2, 1/3, and 1/4 yearly harmonics at 182.62 d, 121.75 d, and 91.31 d that could be produced by the observing schedule. These were visible only in the SuperWASP-North dataset (with the longest time baseline and smallest wrms) and for Gliese 486, as no other SuperWASP comparison star of similar brightness in the same field of view displayed those peaks. A corresponding peak at about 125-130 d appears with false alarm probability (FAP) ≈ 1 % in the generalized Lomb-Scargle periodograms (GLS) (*77*) of ASAS-SN $g'$ and CARMENES VIS H$\alpha$ data (see below). This is consistent with the periods estimated from log $R'_{\text{HK}}$, suggesting the SuperWASP-North peak at ~ 125 d is real. We modelled the SuperWASP and ASAS-SN data using a quasi-periodic Gaussian process (GP) analysis, following (*78*) using the JULIET library (*53*). We used the exp-sin-squared kernel multiplied with a squared-exponential kernel and produced nightly bins for the photometric data. We fitted an offset and a jitter term (in quadrature to the diagonal of the resulting covariance matrix of the GP) and applied distinct GP hyperparameters for the amplitudes for each instrument and photometric band. We also used global GP hyper-parameters for the time scale of the amplitude modulation and the rotation period. This analysis indicated a stellar rotation period $P_{\text{rot,GP}}$ = $130.1^{+1.6}_{-1.2}$ d.

Joint transit and RV analysis

**Tools**

For data and orbital analysis of the Gliese 486 system, we employed the EXO-STRIKER exoplanet toolbox (*21, 79*) to produce a GLS, a maximum likelihood periodogram (MLP) (*80, 81*), transit photometry detrending using the *wōtan* code (*82*), and transit period search using the transit least squares (TLS) package (*83*). For orbital parameter analysis, the EXO-STRIKER offers a fast RV and transit best-fit optimization and sampling schemes such as Markov chain Monte Carlo (MCMC) sampling using the EMCEE sampler (*84*) and the nested sampling technique (*85*) with the DYNESTY sampler (*86*), which were coupled with the CELERITE package (*87*) for GP regression analysis. To build transit light curve models, and extract transit timing variations (TTV), the EXO-STRIKER uses the BAsic Transit Model cAlculatioN package (BATMAN) (*88*).

We also used the JULIET library (*53*) for GP analysis of the ground-based photometry data and for comparison with the EXO-STRIKER analysis.

**Periodogram analysis**

We computed the MLP for period search in RVs and activity indices of Gliese 486. The MLP implementation is similar to a GLS periodogram, but allows for multiple data sets, each with an additive offset and a jitter term (*80*). The log-likelihood (ln $L$) is optimized for each test frequency. Because the MLP fits more parameters, MLP is more computationally expensive than the GLS periodogram, but the MLP is more appropriate for a period search in combined RV data sets that have an unknown variance (that is, RV jitter). We adopted significance thresholds of the likelihood improvements with respect to a model constructed from the same parameters but with zero amplitude, which corresponds to false-alarm probabilities of 10%, 1%, and 0.1%. Fig. S4A shows the MLP periodograms of the CARMENES VIS and NIR, MAROON-X red and blue, HIRES, and HARPS RV time series, separately and combined. The CARMENES VIS and the MAROON-X red and blue data each indicate significant power (FAP < 0.1%) at a period of 1.467 d, much shorter than the stellar rotation period. The MAROON-X data have a short temporal baseline of only ~13.2 d, so the $\Delta\ln L$ power spectrum has lower resolution than the CARMENES, HARPS, and HIRES data. Nevertheless, the MAROON-X data have significant (FAP < 0.1%) power at frequency consistent with the same period. Another strong peak in the CARMENES VIS and MAROON-X periodograms appears at the 1 d alias frequency $f_{alias}$ of the planetary period in the form of $f_{alias} = f_{1d} - f_{planet\ period} \approx 0.31834$ d$^{-1}$ (leading to an alias period of $P_{alias} \approx 3.14$ d), which is no longer seen when the signal of Gliese 486 b is subtracted.

For stars of spectral types M3-4 V, such as Gliese 486, the spectroscopic information (i.e., the number of deep spectral lines) needed for precise RV measurements is not very abundant in the CARMENES NIR spectra (*10, 37*). We find that the 60 CARMENES NIR RVs are less precise and do not have any significant peak with FAP < 0.1% in the MLP periodogram. The HIRES and HARPS data separately do not show significant power with FAP < 0.1% at any frequency either, but the HARPS data set consists of only 12 measurements, while the HIRES dataset consists of 26 measurements with lower precision. The MLP periodogram of the combined data set shows power at 1.467 d, which is dominated by the CARMENES VIS and MAROON-X RVs. The combined data residuals of the joint transit-RV one-planet model (see below) do not show other significant periods.

The MLPs of the CARMENES activity indicators are shown in Fig. S4B. Except for the Hα index, none of them displays signals with significant power of FAP < 0.1% at periods between 1 d and 500 d, in line with previous studies indicating that Gliese 486 is a low-activity star. The Hα MLP has a strong peak at 1/354 d$^{-1}$ and another weaker one, but marginally significant (FAP ~ 1 %), at 1/130 d$^{-1}$. The MLP periodograms of the MAROON-X activity indicators are shown in Fig. S4C. Activity indicators of MAROON-X such as the CRX, Na I D, and Ca IRT do not indicate any significant level of activity in the red and the blue channel. The differential line width and Hα show some marginally significant periodicity (FAP ~ 1 %) in both channels, but without a clear sign of correlation with the RVs over the short MAROON-X temporal baseline.

**Joint modeling fitting**

For the joint fit analysis, we used only data that showed significant RV signal with FAP < 0.1%, or transit light curves consistent with the presence transit events of Gliese 486 b. The used RV datasets were CARMENES VIS and MAROON-X blue and red, whereas we did not use HIRES and CARMENES NIR due their intrinsic large RV scatter and insufficient precision. We found that the HARPS SERVAL RVs generally agree in phase and amplitude with Gliese 486 b, but their overall statistical weight was much smaller than those of CARMENES and MAROON-X, and thus we decided not to include these data in the orbital analysis either. The transit photometry data that we used for the analysis were: TESS Sector 23, the two transit events recorded with MuSCAT2 on 9 May 2020 and 12 May 2020 (hereafter $MuSCAT2_1$ and $MuSCAT2_2$), and the three transit events recorded with LCOGT on 15 May 2020, 24 May 2020, and 5 June 2020 (hereafter $LCOGT_1$, $LCOGT_2$, and $LCOGT_3$). The TJO data and the remaining MuSCAT2 transit data have insufficient precision for precise transit analysis. For increasing the transit signal in the MuSCAT2 data, we combined the four light curves into a single one including *g*, *r*, *i*, and *zs* photometry. All RV and transit data time series were taken in the common time frame of Barycentric Dynamical Time (TDB).

In the first step of our modeling we inspected the PDC TESS light curves. Although the PDC dataset was already corrected for dominant systematics by default, we further corrected it for small systematics, which were still evident in the light curve. In particular, we rejected a dozen obvious outliers and normalized the PDC light curve by fitting a damped stochastically-driven harmonic oscillator (SHO) GP kernel (included in the EXO-STRIKER via CELERITE, 89*)* to capture the non-periodic variation of the light curve. The final product of our detrending was a nearly flat, normalized, TESS light curve, which we adopted to seek for transit signals using TLS. As illustrated in Fig. S5A, we detected a significant TLS signal with false positive rate of < 1% (85), with a period of 1.467 d (as in CARMENES VIS and MAROON-X RV data), together with its harmonics at 0.73 d, 2.93 d, 4.40 d, etc. Fig. S5B shows the TLS power spectrum of the TESS light curve of the joint fit residuals, which have no evidence of additional transit events.

As a second step, using the TESS PDC photometry, we constructed a transit light curve model with planetary orbital parameters: period $P_b$, eccentricity $e_b$, argument of periastron $\omega_b$, inclination $i_b$, time of inferior transit conjunction $t_0$, and the planet semi-major axis and radius $a_b$ and $R_b$ (in units of stellar radius, $R_\star$), respectively. The TESS data parameters adopted in our model were the flux offset and jitter parameters, $TESS_{off}$ and $TESS_{jitt}$. The TESS light curve was detrended simultaneously by the SHO GP model with three hyper-parameters: power $S_0$, characteristic frequency $\omega_0$, and a quality factor of the SHO kernel. We adopted a quadratic limb-darkening model to describe the transit signal shape, adding two more parameters, $u_1$ and $u_2$. We then included the RV model, which added seven additional parameters applied to the RVs. For the CARMENES VIS and MAROON-X red and blue datasets we fitted for the RV offsets, RV jitters, and the RV signal semi-amplitude *K*, which constrains the planetary mass.

The rest of orbital parameters are common for the transit and RV model components. In total, the joint model has 21 data and orbital free parameters.

As an alternative analysis, we built a more complex joint model including the MuSCAT2 and LCOGT photometry. For modeling the TESS, MuSCAT2, and LCOGT light curves together with the RVs from CARMENES VIS and MAROON-X red and blue, we adopted different quadratic limb-darkening models and optimized the quadratic limb-darkening parameters for each instrument with six parameters: TESS $u_1$ and $u_2$, MuSCAT2$_{1,2}$ $u_1$ and $u_2$, and LCOGT$_{1,2,3}$ $u_1$ and $u_2$. The ground-based transit MuSCAT2 and LCOGT data were simultaneously detrended with a linear model against airmass at the time of measurement, thus adding five more parameters. We also varied the flux offset and jitter parameter of each transit light curve data separately, which translated into six offset and six jitter transit data parameters. In total, this alternative model has 40 free parameters.

For the modelling fitting process, we adopted a dynamical nested sampling with DYNESTY, with 100% weight on the posterior convergence (86). For all parameters we adopted priors, which are summarized in Table S5. Our nested sampling test represented a forced, high-density, multi-dimensional parameter volume search, the posterior estimates of which were adopted as our final results. The parameter posterior estimates of the two joint models described above (hereafter CMT, for the model including CARMENES VIS, MAROON-X and TESS, and CMT+LM, for the model that adds LCOGT and MuSCAT2) are summarized in Table S6. Fig. 1 shows the phase-folded CMT data and model, Fig. S6 shows the TESS, MuSCAT2$_{1,2}$, and LCOGT$_{1,2,3}$ flux time series and the transit light curve component of the CMT+LM model, and Fig. S7 shows the detrended phase-folded data of the CMT+LM model. The posterior distributions of the nested sampling parameters of both models are shown in the corner plots of Figs. S8 and S9, respectively. Both models are consistent with each other within the estimated uncertainties, although the CMT+LM model has larger parameter uncertainties. We attribute this to the much larger parameter space (21 versus 40 parameters), which produces additional covariance with the orbiting parameters. The noisier MuSCAT2 and LCOGT data with respect to TESS do not contribute substantial information to the orbital and physical determination of Gliese 486 b. Therefore, in Table 1 and the reminder of our analysis we report only the parameters obtained from the CMT model.

The orbital eccentricity of Gliese 486 b is not constrained. Our full-Keplerian modelling was done with free $e_b$, $\omega_b$, or $e_b\sin(\omega_b)$, $e_b\cos(\omega_b)$ parameterization, and both solutions provided only an upper limit on the eccentricity of $e_b < 0.05$ at the 68.3% confidence level. A forced circular model of Gliese 486 b with $e_b$ fixed at 0 (but $t_0$ varied to assure transit event at $t_0 \sim$ 2458931.16) led to solutions which are statistically indistinguishable from the full Keplerian model. The CMT model has a Bayesian log-evidence of ln $Z$ = 76406.2 ± 0.4 for the circular model and ln $Z$ = 76405.1 ± 0.4 for the full-Keplerian model. The CMT+LM model is similar: ln $Z$ = 84642.6 ± 0.4 for the circular model and ln $Z$ = 84641.7 ± 0.4 for the full-Keplerian model. This low orbital eccentricity is what we expect given the planet's proximity to the star, which should cause tidal circularization. We investigated the star-planet tides of the Gliese 486 system using the EQTIDE code (22), which calculates the tidal evolution of two bodies based on standard models (89, 90, 91). For Gliese 486 b we adopted the Earth's value $k_2/Q$ = 0.025 from (92) and initial planetary rotational period of 0.5 d, whereas for the star Gliese 486 we adopted $k_2/Q = 2 \cdot 10^6$ and an initial stellar rotational period of 130 d. Fig. S10 shows the eccentricity decay due to star-planet tides from our tidal evolution simulations. We tried a set of different initial semi-major axes and eccentricities a few percent larger than the observed, and found that Gliese 486 b reached synchronous rotation within < 10 000 yr and that, on average, its planetary orbit was fully circularized in only ~ one million years. For our final orbital solution of Gliese 486 b, we therefore adopted the simpler circular orbit model. Our final orbital solution for Gliese 486 b is given in Table 1.

The CARMENES VIS data show small residual scatter of wrms = 1.87 m s$^{-1}$ and an RV jitter level of 1.45 m s$^{-1}$. MAROON-X blue channel data show wrms = 1.12 m s$^{-1}$ and an RV jitter level of 0.70 ms$^{-1}$, while the red channel shows wrms = 0.42 m s$^{-1}$ and an RV jitter level of only 0.25 m s$^{-1}$. The MAROON-X red radial velocities have the lowest scatter ever seen for an M dwarf without applying corrections for activity-induced jitter.

Search for transit timing variations

To search for possible TTVs, we performed two independent analyses including all detected transit data available. The first analysis was done using the EXO-STRIKER by adopting the CMT+LM model, but allowing for variable transit mid times. In this model, the orbital period $P_b$ was fixed at its best-fitting value, while the transit times $t_0$ to $t_{52}$ were allowed to vary (but only fitting the 18 individual times-of-transits for which we had data), thus adding 16 more fitting parameters to the base model. The second test was done with JULIET, which was applied only to the transit data. In this scheme, all the transit parameters across each individual TESS, MuSCAT2, and LCOGT transit were shared, except for the limb-darkening coefficients (which were individual to each instrument), the 18 individual times-of-transits, out-of-transit fluxes, and the coefficients of linear models in airmass, which were used to detrend each of the ground-based light curves simultaneously in the modelling procedure.

We detected some marginal TTVs in the order of a few minutes in the *TESS* data, and larger variations on the LCOGT transits, but with higher TTV uncertainty. Using the EXO-STRIKER and JULIET we qualitatively compared a fit using a linear ephemeris (that is, non-TTV model) and a model that allows TTVs. We found a very strong Bayesian evidence in favor of a linear ephemeris i.e., no significant TTVs arising from the combined transit photometry (Δln Z ~ 44 in the case the EXO-STRIKER, Δln Z ~ 37 in the case of JULIET). We also used the EXO-STRIKER to dynamically model the extracted TTVs, but we could not explain these variations by another non-transiting planet perturbing Gliese 486 b. This is consistent with the RV data, which did not show any evidence for another planet. We conclude that there is no reason to prefer TTVs over linear ephemeris and evidence of only a single planet Gliese 486 b.

**Supplementary Text**

Prospects for atmospheric investigation of Gliese 486 b

Fig. 3A shows the expected transmission signal of the planetary atmospheres of all known rocky planets (with $R_p$ between 0.5 and 2.0 $R_E$) with measured masses and radii that transit M dwarfs as a function of the host star magnitude in the *K*s band. In all cases, a mean molecular weight $\mu = 18$ for a water (steam)-dominated atmosphere was assumed. Higher transmission signal values around bright stellar host magnitudes provide more favorable conditions for detecting a possible atmosphere, while planets with lower transmission signals around faint stars are more technically challenging to characterize. Three target sub-groups are apparent. The first is rocky planets transiting around very bright host stars, visible with the naked eye from dark sites. These are generally G- and K-type main-sequence stars, and the prospects for their atmospheric investigation and characterization are higher because of the host star brightness. Members of this group are 55 Cnc e (*30*), HD 219134 b and c (*93*), and BD-02 5958 b and c (*94*) (π Men c (*95*), with a density of about 2.8 $10^3$ kg m$^{-3}$, is not a rocky planet). However, except for the poorly understood variability of 55 Cnc e (*96*), none has a detected atmosphere. The second group are planets orbiting M-dwarf hosts that have better prospects for atmospheric detection, as the small size of the host star compensates for their much dimmer brightness. In this group, the largest atmospheric signals are expected for the TRAPPIST-1 planets because of the high radius ratio between the planets and the host star. Gliese 486 b is also favorable for rocky planet atmosphere searches. Gliese 486 b is similar to GJ 357 b (*97*) in terms of planet parameters and prospects for atmospheric investigation. These planets have similar suitability: the known super-Earths around non-M stars, Gliese 486 b, and the TRAPPIST-1 system. A continuously updated compendium of transiting planets with measured mass around M dwarfs is available in (98).

The combination of its small radius and high equilibrium temperature makes Gliese 486 b unlikely to have retained a large atmosphere. With a radius of about 1.3 $R_E$, we expect Gliese 486 b to have lost its primordial hydrogen-helium atmosphere due to photoevaporation processes (*29, 99, 100*). At the current planet location the atmosphere could have been lost during the earlier phases of Gliese 486 stellar evolution. However, whether rocky planets around M dwarfs are able to retain a substantial fraction of their atmospheres and, if so, at which ranges of mass and $T_{eq}$ remains an open question. Gliese 486 b could be used to test these mechanisms.

At present, LHS 3844 b, a 1.3 $R_E$ planet around an M5 V star, is the most thoroughly investigated small rocky planet in search for an atmosphere. Its thermal phase curve has been searched for signs of atmospheric heat redistribution (101). Those authors determined that the data were best explained by a bare rock model with a low Bond albedo, supporting theoretical predictions that hot terrestrial planets orbiting small stars may not retain substantial atmospheres (*99, 102*). However, LHS 3844 b has an orbital period 3.2 times shorter than Gliese 486 b and $T_{eq}$ hotter by 100 K. LHS 3844 b does not have a measured mass, limiting interpretation of its atmosphere. The brightness of the host star makes Gliese 486 b a more suitable target for phase curve characterization and epoch of superior transit conjunction (secondary eclipse time) determination and, thus, determining the day and night side temperatures of the planet. Our joint model with free planet eccentricity constrains the secondary eclipse time to $2458931.88643^{+0.00769}_{-0.00829}$ d, suitable for scheduling future observations.

**Table S1. Radial velocity time-series from the CARMENES VIS channel spectra.** Only a subset of the data analyzed in this paper is shown here. A machine-readable version of the full dataset, including the spectroscopic activity indices, is available in Data S1.

| Barycentric Julian Date, BJD | Radial velocity, RV (m s$^{-1}$) | Radial velocity uncertainty, σRV (m s$^{-1}$) |
|---|---|---|
| 2457400.74081 | 4.52 | 1.07 |
| 2457401.74239 | 0.07 | 1.30 |
| 2457418.71847 | -2.32 | 1.14 |
| 2457421.70507 | -2.65 | 0.98 |
| 2457426.69298 | 0.91 | 1.10 |
| 2457442.60293 | -2.97 | 0.91 |
| 2457442.62657 | -3.46 | 0.93 |
| 2457476.51979 | -2.88 | 1.35 |
| 2457492.53441 | -1.28 | 1.62 |

**Table S2. Radial velocity time-series from the CARMENES NIR channel spectra.** Only a subset of the data analyzed in this paper is shown here. A machine-readable version of the full dataset, including the spectroscopic activity indices, is available in Data S1.

| Barycentric Julian Date, BJD | Radial velocity, RV (m s$^{-1}$) | Radial velocity uncertainty, σRV (m s$^{-1}$) |
|---|---|---|
| 2457788.52216 | -24.63 | 10.92 |
| 2457802.65175 | -4.70 | 5.12 |
| 2457856.53224 | -19.15 | 4.22 |
| 2457876.53529 | -16.20 | 4.48 |
| 2457896.4259 | -15.63 | 3.89 |
| 2457950.37141 | -17.72 | 9.92 |
| 2458122.69387 | -12.63 | 3.94 |
| 2458141.58966 | -14.78 | 4.94 |
| 2458206.57208 | -11.17 | 5.61 |

**Table S3. Radial velocity time-series from the MAROON-X red channel spectra.** Only a subset of the data analyzed in this paper is shown here. A machine-readable version of the full dataset, including the spectroscopic activity indices, is available in Data S1.

| Barycentric Julian Date, BJD | Radial velocity, RV (m s$^{-1}$) | Radial velocity uncertainty, σRV (m s$^{-1}$) |
|---|---|---|
| 2458989.74702 | 1.74 | 0.46 |
| 2458989.75182 | 1.37 | 0.40 |
| 2458991.82562 | -2.38 | 0.40 |
| 2458991.83039 | -1.98 | 0.34 |
| 2458992.85416 | -0.58 | 0.27 |
| 2458992.85888 | -1.02 | 0.40 |
| 2458993.82807 | 4.14 | 0.30 |
| 2458993.83285 | 4.17 | 0.46 |
| 2458994.77985 | -2.26 | 0.72 |

**Table S4. Radial velocity time-series from the MAROON-X blue channel spectra.** Only a subset of the data analyzed in this paper is shown here. A machine-readable version of the full dataset, including the spectroscopic activity indices, is available in Data S1.

| Barycentric Julian Date, BJD | Radial velocity, RV (m s$^{-1}$) | Radial velocity uncertainty, σRV (m s$^{-1}$) |
|---|---|---|
| 2458989.74701 | 3.30 | 1.01 |
| 2458989.75179 | 1.72 | 0.86 |
| 2458991.82561 | -2.75 | 0.77 |
| 2458991.83037 | -3.37 | 0.84 |
| 2458992.85415 | -1.55 | 0.87 |
| 2458992.85885 | -2.59 | 0.82 |
| 2458993.82806 | 4.19 | 1.13 |
| 2458993.83284 | 2.99 | 1.01 |
| 2458994.77982 | -2.56 | 1.68 |

**Table S5. Adopted parameter priors.** These prior probabilities were used as input to the modeling of photometry (TESS, MuSCAT2, LCOGT) and radial velocities (CARMENES VIS, MAROON-X red, MAROON-X blue). The notations of *N*, *U*, and *J* represent normal, uniform, and Jeffrey's prior probability distributions.

| Parameter | Adopted priors |
|---|---|
| $K_b$ (m s$^{-1}$) | $U(0.01, 5.00)$ |
| $P_b$ (d) | $U(1.46500, 1.47500)$ |
| $e_b$ | $U(0.0, 0.3)$, or fixed at 0 |
| $\omega_b$ (deg) | $U(0.0, 360.0)$, or undefined when $e_b = 0$ |
| $e_b \sin(\omega_b)$ | $U(-1.0, 1.0)$ |
| $e_b \cos(\omega_b)$ | $U(-1.0, 1.0)$ |
| $i_b$ (deg) | $U(85.00, 95.00)$ |
| $t_0 - 2450000$ (BJD) | $U(8931.04, 8931.26)$ |
| $a_b/R_\star$ | $U(5.00, 15.00)$ |
| $R_b/R_\star$ | $U(0.01, 0.05)$ |
| RV offset CARMENES (m s$^{-1}$) | $U(-5.00, 5.00)$ |
| RV jitter CARMENES (m s$^{-1}$) | $J(0.01, 5.00)$ |
| RV offset MAROON-X red (m s$^{-1}$) | $U(-5.00, 5.00)$ |
| RV jitter MAROON-X red (m s$^{-1}$) | $J(0.01, 5.00)$ |
| RV offset MAROON-X blue (m s$^{-1}$) | $U(-5.00, 5.00)$ |
| RV jitter MAROON-X blue (m s$^{-1}$) | $J(0.01, 5.00)$ |
| Transit offset TESS (ppm) | $N(0.0, 1000.0)$ |
| Transit jitter TESS (ppm) | $J(1.0, 3000)$ |

| Transit offset MuSCAT2$_{1,2}$ (ppm) | $N(0.0, 1000.0)$ |
|---|---|
| Transit jitter MuSCAT2$_{1,2}$ (ppm) | $J(1.0, 3000)$ |
| Transit offset LCOGT$_{1,2,3}$ (ppm) | $N(0.0, 1000.0)$ |
| Transit jitter LCOGT$_{1,2,3}$ (ppm) | $J(1.0, 3000)$ |
| TESS GP SHO $S_0$ | $J(0.0001, 0.0100)$ |
| TESS GP SHO Q | $J(0.0001, 0.5000)$ |
| TESS GP SHO $\omega_0$ | $J(0.0001, 2.0000)$ |
| Linear detrend. coef. MuSCAT2$_{1,2}$ and LCOGT$_{1,2,3}$ | $U(-0.1, 0.1)$ |
| Quad. limb-dark. TESS $u_1$ | $U(0.00, 1.00)$ |
| Quad. limb-dark. TESS $u_2$ | $U(0.00, 1.00)$ |
| Quad. limb-dark. MuSCAT2 $u_1$ | $U(0.00, 1.00)$ |
| Quad. limb-dark. MuSCAT2 $u_2$ | $U(0.00, 1.00)$ |
| Quad. limb-dark. LCOGT $u_1$ | $U(0.00, 1.00)$ |
| Quad. limb-dark. LCOGT $u_2$ | $U(0.00, 1.00)$ |

**Table S6. Results of the joint fit model fitting.** Best fitting values and uncertainties are listed as extracted from the posterior probability distributions of the CMT and CMT+ML models (Figures S8 and S9, respectively).

| Parameter | CMT+ML fit | CMT fit |
|---|---|---|
| $K_p$ [m s$^{-1}$] | $3.358^{+0.099}_{-0.164}$ | $3.371^{+0.070}_{-0.081}$ |
| $P_p$ [d] | $1.467111^{+0.000050}_{-0.000026}$ | $1.467119^{+0.000031}_{-0.000030}$ |
| $i_p$ [deg] | $88.6^{+1.0}_{-1.4}$ | $88.4^{+1.1}_{-1.4}$ |
| $t_0$ [d] | $2458931.15939^{+0.00056}_{-0.00067}$ | $2458931.15935^{+0.00042}_{-0.00042}$ |
| $a_p/R_\star$ | $10.94^{+0.55}_{-1.22}$ | $10.80^{+0.57}_{-1.02}$ |
| $R_p/R_\star$ | $0.0366^{+0.0011}_{-0.0026}$ | $0.0365^{+0.0011}_{-0.0014}$ |
| RV off. CARMENES−VIS [m s$^{-1}$] | $-0.15^{+0.31}_{-0.32}$ | $-0.19^{+0.22}_{-0.23}$ |
| RV off. MAROON−X red [m s$^{-1}$] | $0.105^{+0.084}_{-0.089}$ | $0.111^{+0.057}_{-0.055}$ |
| RV off. MAROON−X blue [m s$^{-1}$] | $0.09^{+0.20}_{-0.21}$ | $0.10^{+0.14}_{-0.13}$ |
| RV jitter CARMENES−VIS [m s$^{-1}$] | $1.42^{+0.26}_{-0.37}$ | $1.47^{+0.22}_{-0.20}$ |
| RV jitter MAROON−X red [m s$^{-1}$] | $0.258^{+0.163}_{-0.080}$ | $0.245^{+0.071}_{-0.066}$ |
| RV jitter MAROON−X blue [m s$^{-1}$] | $0.65^{+0.24}_{-0.30}$ | $0.67^{+0.18}_{-0.19}$ |
| Transit offset *TESS* [ppm] | $0^{+1900}_{-1800}$ | $60^{+850}_{-890}$ |
| Transit offset MuSCAT2$_1$ [ppm] | $-200^{+2000}_{-2100}$ | ... |
| Transit offset MuSCAT2$_2$ [ppm] | $1300^{+1800}_{-1500}$ | ... |
| Transit offset LCOGT$_1$ [ppm] | $-2300^{+940}_{-1210}$ | ... |
| Transit offset LCOGT$_2$ [ppm] | $800^{+1900}_{-1700}$ | ... |
| Transit offset LCOGT$_3$ [ppm] | $-3950^{+770}_{-1120}$ | ... |
| Transit jitter *TESS* [ppm] | $4.6^{+11.5}_{-2.9}$ | $4.0^{+6.4}_{-2.3}$ |
| Transit jitter MuSCAT2$_1$ [ppm] | $24^{+146}_{-21}$ | ... |
| Transit jitter MuSCAT2$_2$ [ppm] | $20^{+107}_{-17}$ | ... |
| Transit jitter LCOGT$_1$ [ppm] | $1790^{+210}_{-1250}$ | ... |
| Transit jitter LCOGT$_2$ [ppm] | $34^{+285}_{-31}$ | ... |
| Transit jitter LCOGT$_3$ [ppm] | $930^{+170}_{-710}$ | ... |
| *TESS* GP-SHO $S_0$ | $0.00102^{+0.00319}_{-0.00078}$ | $0.00074^{+0.00320}_{-0.00055}$ |
| *TESS* GP-SHO Q | $0.0108^{+0.0197}_{-0.0077}$ | $0.0093^{+0.0131}_{-0.0063}$ |
| *TESS* GP-SHO $\omega_0$ | $0.27^{+0.46}_{-0.18}$ | $0.36^{+0.56}_{-0.20}$ |
| Linear trend MuSCAT2$_1$ | $0.0014^{+0.0016}_{-0.0015}$ | ... |
| Linear trend MuSCAT2$_2$ | $0.0038^{+0.0014}_{-0.0016}$ | ... |
| Linear trend LCOGT$_1$ | $-0.00148^{+0.00070}_{-0.00058}$ | ... |
| Linear trend LCOGT$_2$ | $0.0006^{+0.0012}_{-0.0014}$ | ... |
| Linear trend LCOGT$_3$ | $-0.00252^{+0.00086}_{-0.00071}$ | ... |
| $u_1$ TESS | $0.29^{+0.25}_{-0.18}$ | $0.26^{+0.21}_{-0.16}$ |
| $u_2$ TESS | $0.39^{+0.30}_{-0.24}$ | $0.42^{+0.31}_{-0.26}$ |
| $u_1$ MuSCAT2 | $0.48^{+0.25}_{-0.27}$ | ... |
| $u_2$ MuSCAT2 | $0.52^{+0.29}_{-0.31}$ | ... |
| $u_1$ LCOGT | $0.51^{+0.27}_{-0.28}$ | ... |
| $u_2$ LCOGT | $0.48^{+0.31}_{-0.29}$ | ... |
| $M_p$ [M$_\oplus$] | $2.80^{+0.14}_{-0.19}$ | $2.82^{+0.11}_{-0.12}$ |
| $a_p$ [au] | $0.01734^{+0.00026}_{-0.00027}$ | $0.01734^{+0.00026}_{-0.00027}$ |
| $R_p$ [$R_\oplus$] | $1.305^{+0.068}_{-0.107}$ | $1.305^{+0.063}_{-0.067}$ |
| $T_{eq}$ [K] | $701^{+13}_{-13}$ | $701^{+13}_{-13}$ |
| $S$ [$S_\oplus$] | $40.3^{+1.5}_{-1.4}$ | $40.2^{+1.5}_{-1.4}$ |
| $g$ [m s$^{-2}$] | $16.1^{+2.6}_{-1.8}$ | $16.2^{+1.9}_{-1.6}$ |
| $\rho_b$ [$10^3$ kg m$^{-3}$] | $6.9^{+1.7}_{-1.1}$ | $7.0^{+1.2}_{-1.0}$ |
| $v_{esc}$ [km s$^{-1}$] | $16.37^{+0.70}_{-0.64}$ | $16.44^{+0.55}_{-0.52}$ |
| Impact parameter $b$ | $0.27^{+0.21}_{-0.18}$ | $0.29^{+0.20}_{-0.20}$ |
| Transit duration [h] | $1.021^{+0.046}_{-0.027}$ | $1.025^{+0.031}_{-0.023}$ |
| $\rho_\star$ [$10^3$ kg m$^{-3}$] | $11.5^{+1.8}_{-3.4}$ | $11.1^{+1.9}_{-2.8}$ |

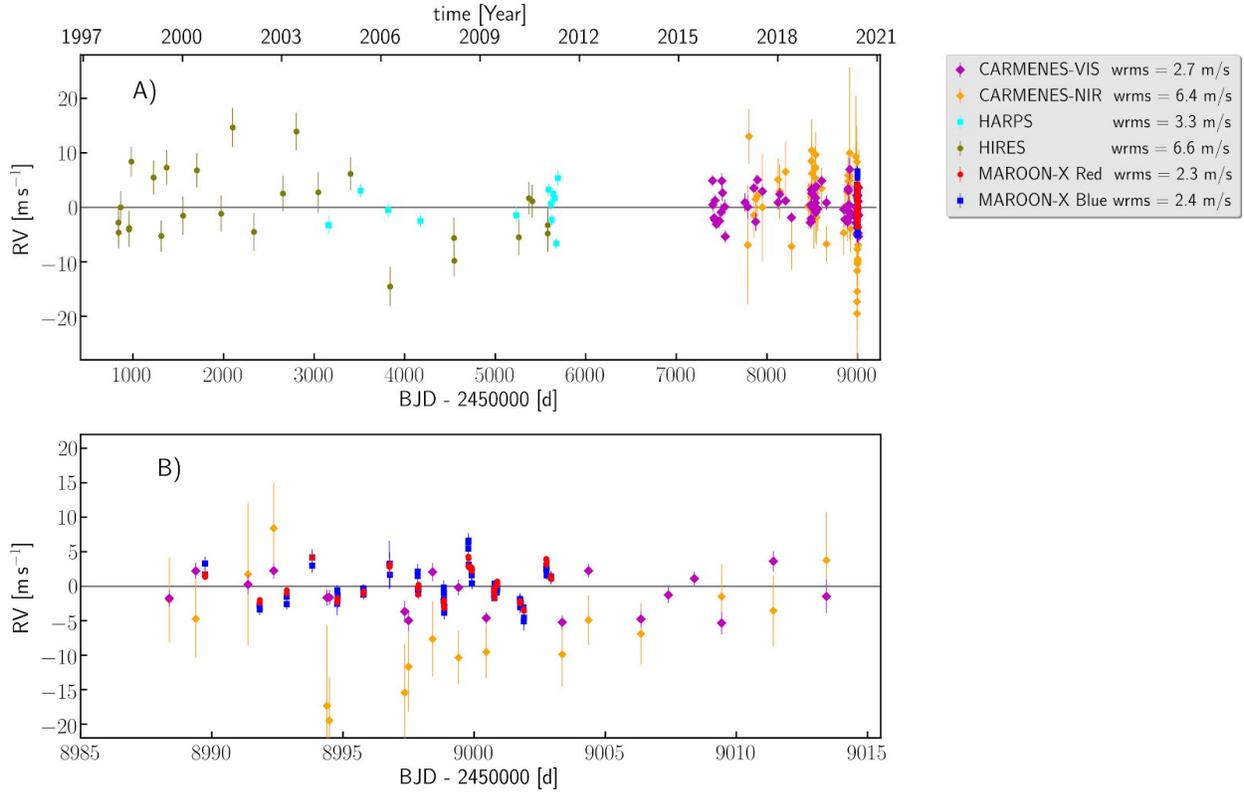

**Fig. S1. RV data for Gliese 486.** Panel **(A)** shows 27 HIRES RVs (green circles), 12 HARPS RVs (cyan squares), 76 CARMENES VIS RVs (magenta diamonds), 60 CARMENES NIR (amber diamonds), and 65 MAROON-X blue (blue squares) and red (red circles) RVs. The data error bars indicate the 1σ uncertainties of the measurements. The time baseline of the observations is from January 1998 to May 2020. A HIRES outlier at BJD = 2452006.986 (RV ~ −38 m s$^{-1}$) falls outside of the plotting range. Calendar years are indicated at the top for reference. Panel **(B)** shows a zoomed baseline between BJD = 2458985 and 2459015 when high-cadence RVs were obtained.

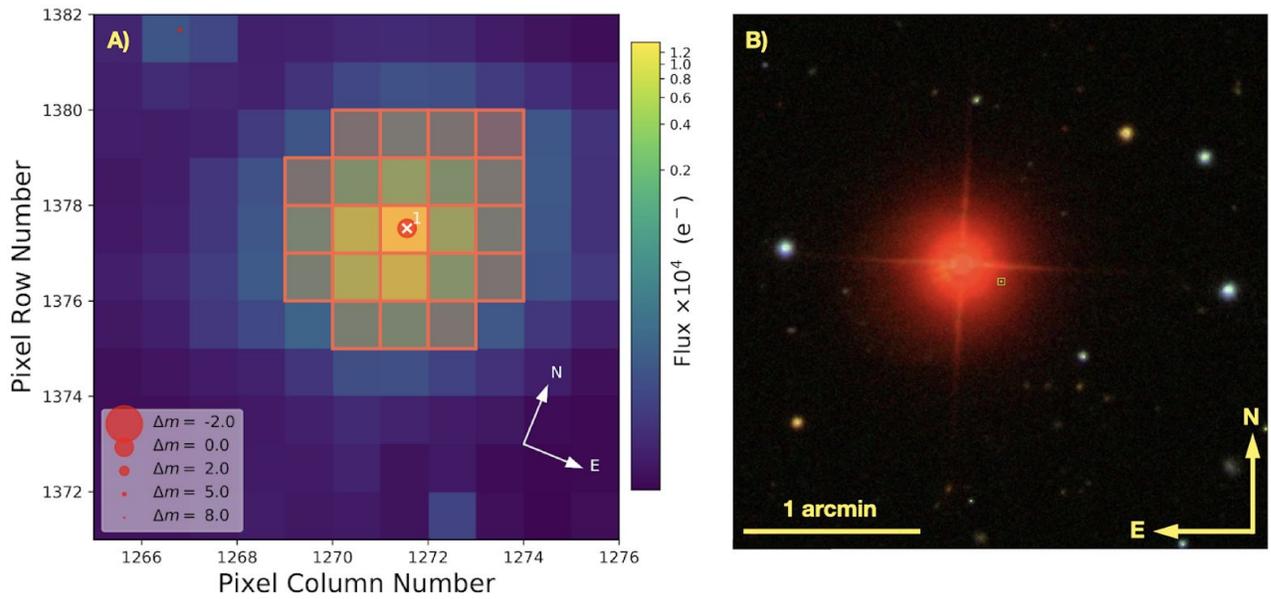

**Fig. S2.** *TESS* **Sector 23 TPF and a false-color, 3×3 arcmin² SDSS9 image of Gliese 486. (A)** The TPF electron counts are color-coded by flux, the orange bordered pixels are used in SAP, and the scale is 21 arcsec pixel$^{-1}$. **(B)** A green square in the *g'r'i'* SDSS9 (*52*) composition (epoch of observation: J2003.32) marks the location of the star in early 2020. In both fields of view, Gliese 486 is the brightest star.

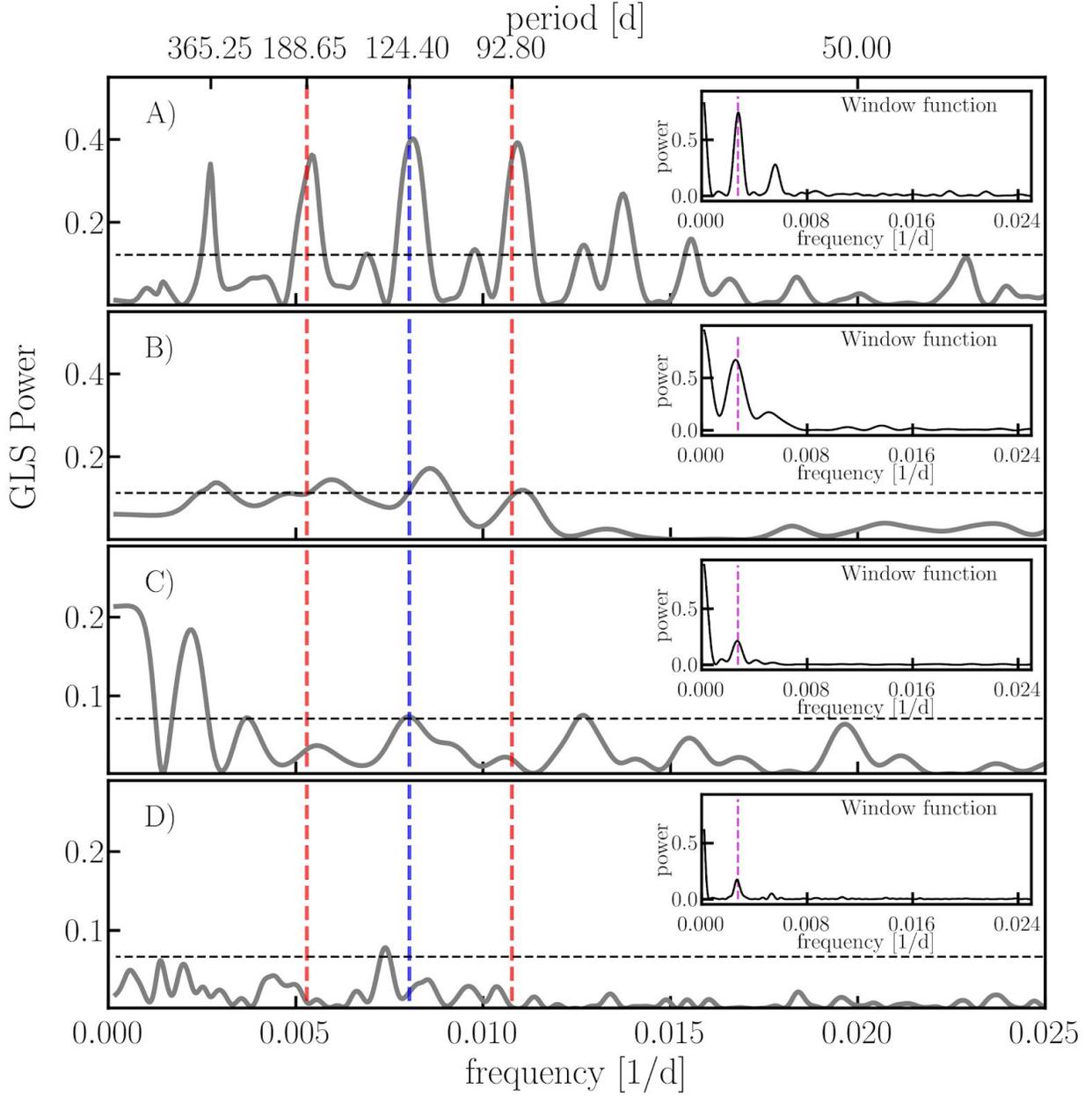

**Fig. S3. GLS power spectrum of the photometric data from SuperWASP and ASAS-SN of Gliese 486.** **(A)** SuperWASP North, **(B)** SuperWASP South, **(C)** ASAS-SN *g'* and **(D)** ASAS-SN *V* band ground-based photometry. The inset panels show the discrete Fourier transform window function of the observations. The blue vertical dashed line indicates a peak that is close to the most likely stellar rotational period of Gliese 486 obtained from GP ($P_{rot} \sim 130$ d), while the red vertical dashed lines indicate the first two one-year aliases of this signal.

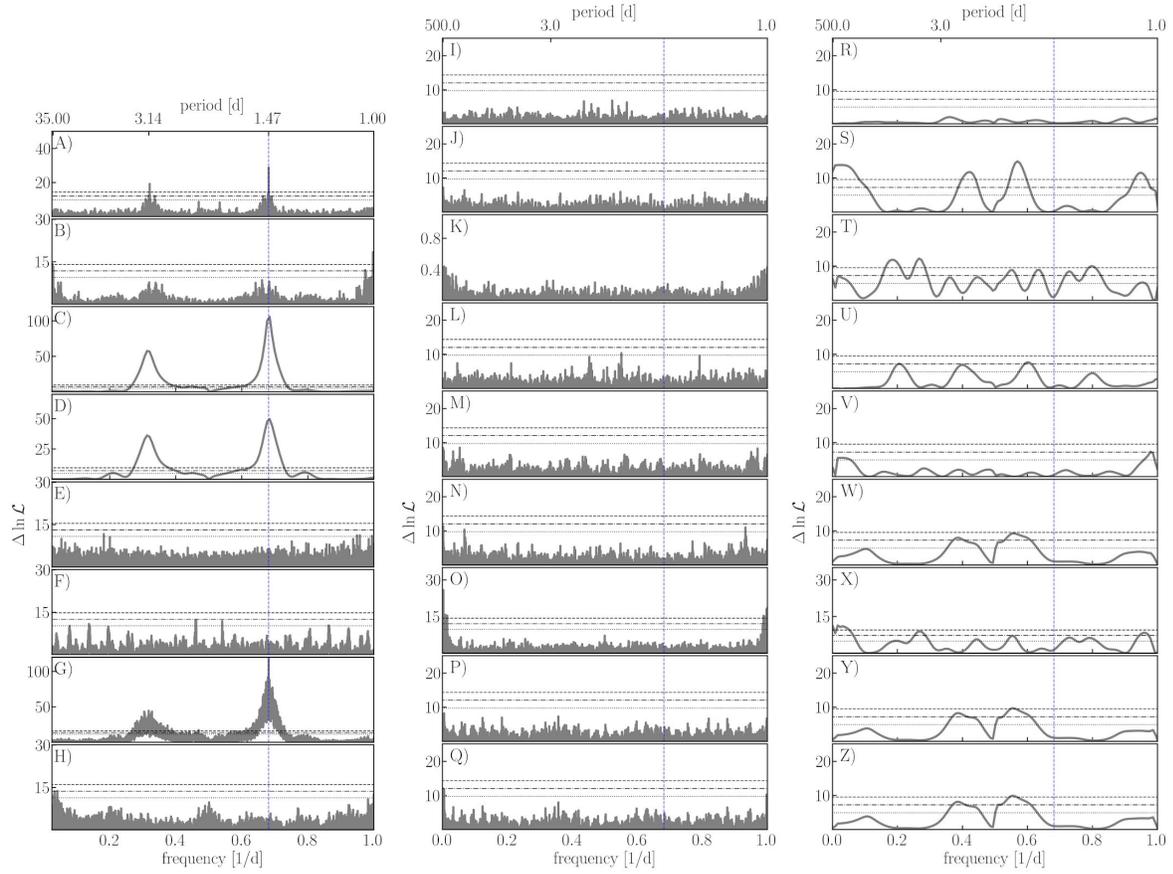

**Fig. S4. Maximum logarithmic likelihood periodograms of the spectroscopic data of Gliese 486.** Left panels are **(A)** CARMENES VIS RVs, **(B)** CARMENES NIR RVs, **(C)** MAROON-X red RVs, **(D)** MAROON-X blue RVs, **(E)** HIRES RVs, **(F)** HARPS, **(G)** all RVs together, **(H)** best-fit residuals of all RVs; middle panels are **(I)** CARMENES-VIS BIS, **(J)** CARMENES-VIS CON, **(K)** CARMENES-VIS FWHM, **(L)** CARMENES-VIS Ca IRT, **(M)** CARMENES-VIS CRX, **(N)** CARMENES-VIS dLW, **(O)** CARMENES-VIS Hα, **(P)** CARMENES-VIS Na D1, **(Q)** CARMENES-VIS Na D2; right panels are **(R)** MAROON-X red CRX, **(S)** MAROON-X red dLW, **(T)** MAROON-X red Hα, **(U)** MAROON-X red Ca IRT, **(V)** MAROON-X blue CRX, **(W)** MAROON-X blue dLW, **(X)** MAROON-X blue Hα, **(Y)** MAROON-X Na D1, **(Z)** MAROON-X Na D2. Panels **(A)-(H)** show only the period range of 1-40 d (no significant lower frequency signals are detected in the RV data). The orbital frequency of Gliese 486 b is $P_\mathrm{b}$ = 1.467 d (blue dashed vertical line) is apparent in the CARMENES VIS and MAROON-X data. The second strongest peak at ~3.14 d is the 1 d alias frequency. Horizontal lines indicate the Δln $L$ significance levels that correspond to FAP = 10% (dotted), 1% (dot-dashed), and 0.1% (dashed).

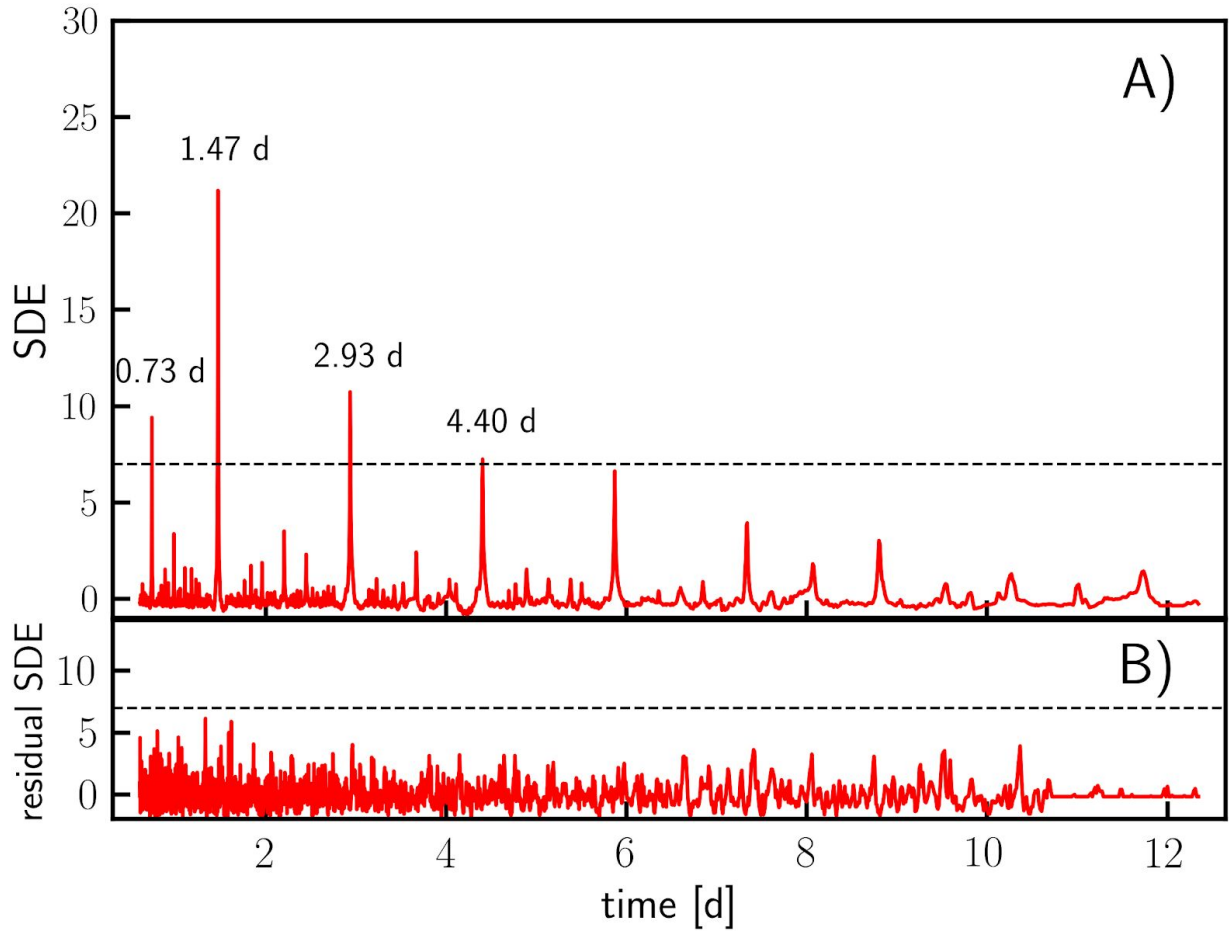

**Fig. S5. TLS power spectra of the detrended TESS Sector 23 PDC light curve of Gliese 486.** **(A)** The planetary transit signal at $P_b = 1.467$ d is accompanied by harmonics at 0.73, 2.93, and 4.40 d. **(B)** TESS residuals of the one-planet transit model. The horizontal dashed line indicates the signal detection efficiency (SDE) power level of 7.0, which corresponds to a TLS false positive rate of 1 % (85).

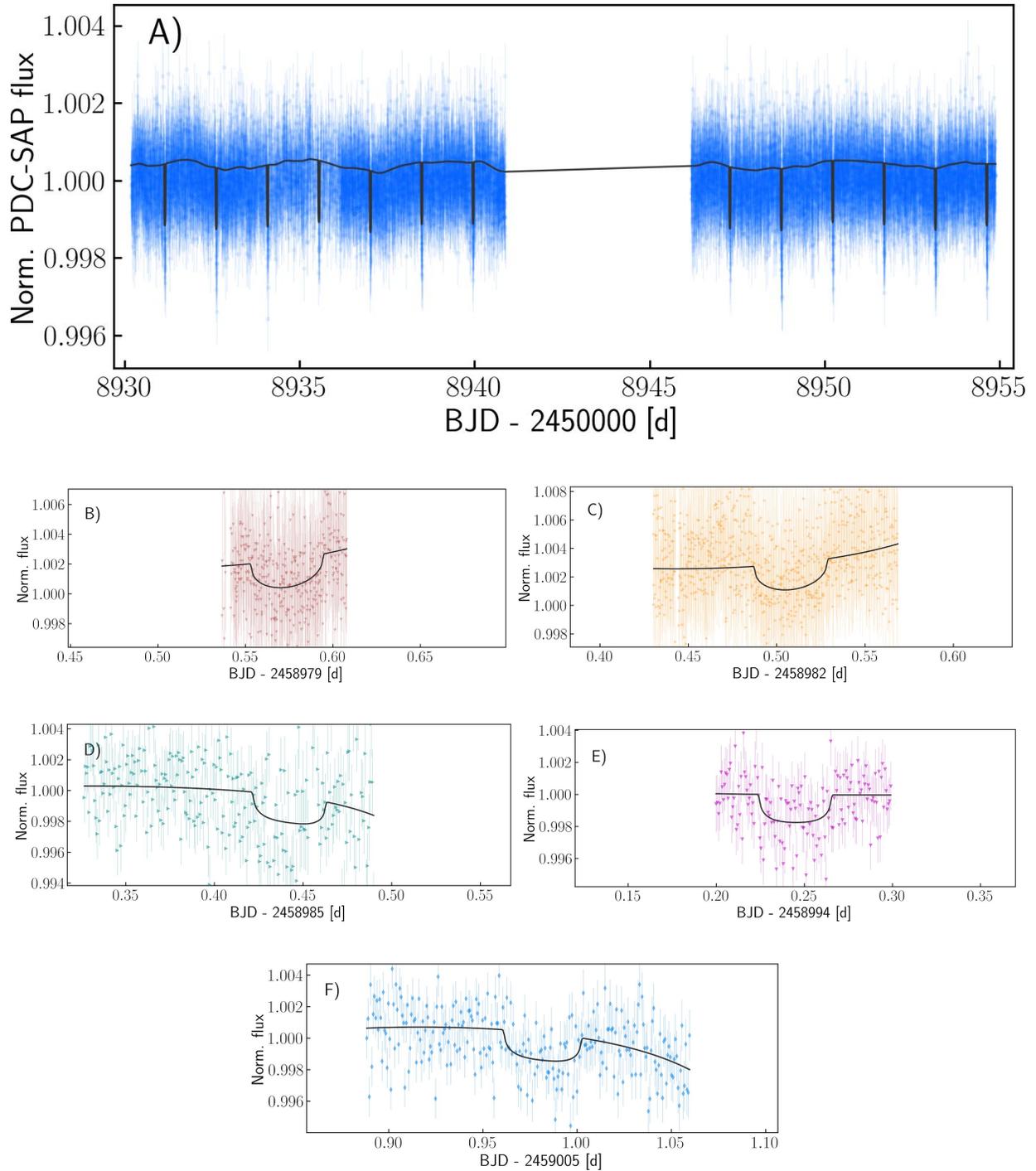

**Fig. S6. Transit photometry of Gliese 486.** The transit component of the joint model is shown with a black solid line. (**A**) PDC data from Sector 23 of *TESS*. (**B**) Ground based data of Gliese 486 from MuSCAT2$_1$ and (**C**) MuSCAT2$_2$. (**D**) Ground based data of Gliese 486 from LCOGT$_1$, (**E**) LCOGT$_2$, and (**F**) LCOGT$_3$. Error bars indicate 1σ uncertainties of individual measurements.

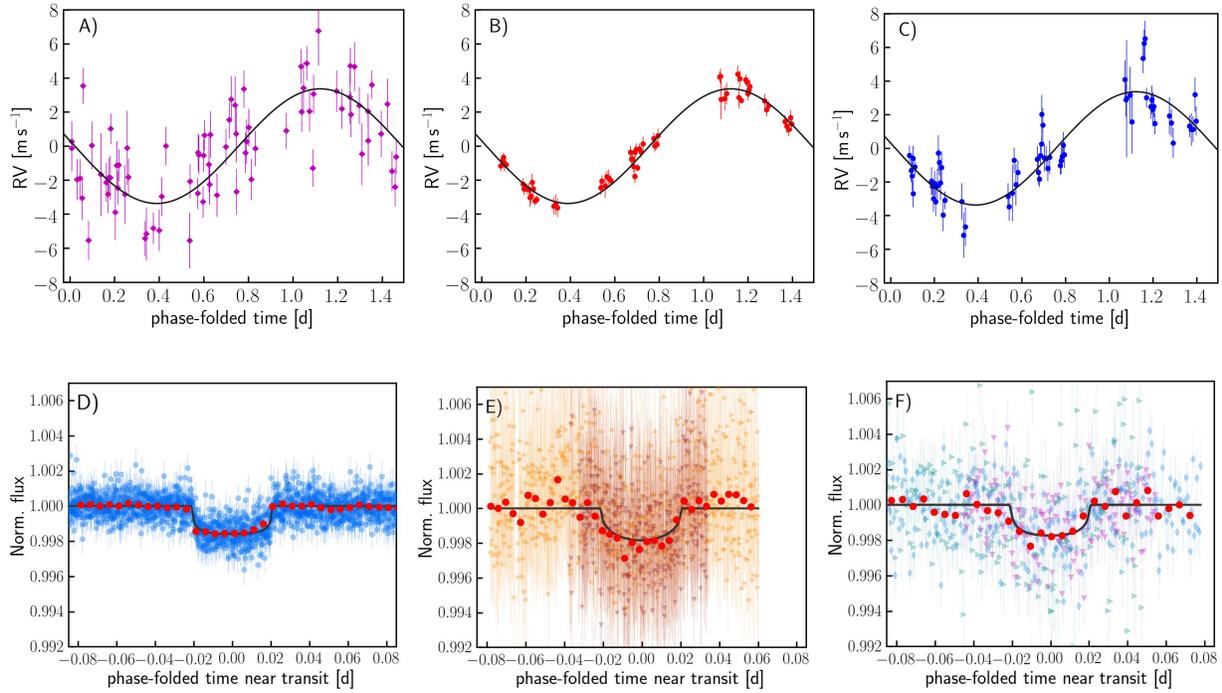

**Fig. S7. Same as Fig. 1, but for the CMT+LM model and datasets.** Phase-folded CARMENES VIS **(A)**, MAROON-X red **(B)**, and MAROON-X blue RV data **(C)**. Phase-folded sector 23 *TESS* data **(D)**, MuSCAT2 data obtained on two nights (9 May 2020: amber, 12 May 2020: brown) **(E)**, and LCOGT data obtained on three nights (15 May 2020: cyan, 24 May 2020: magenta, 5 June 2020: light blue) **(F)**. Error bars indicate 1σ uncertainties of individual measurements.

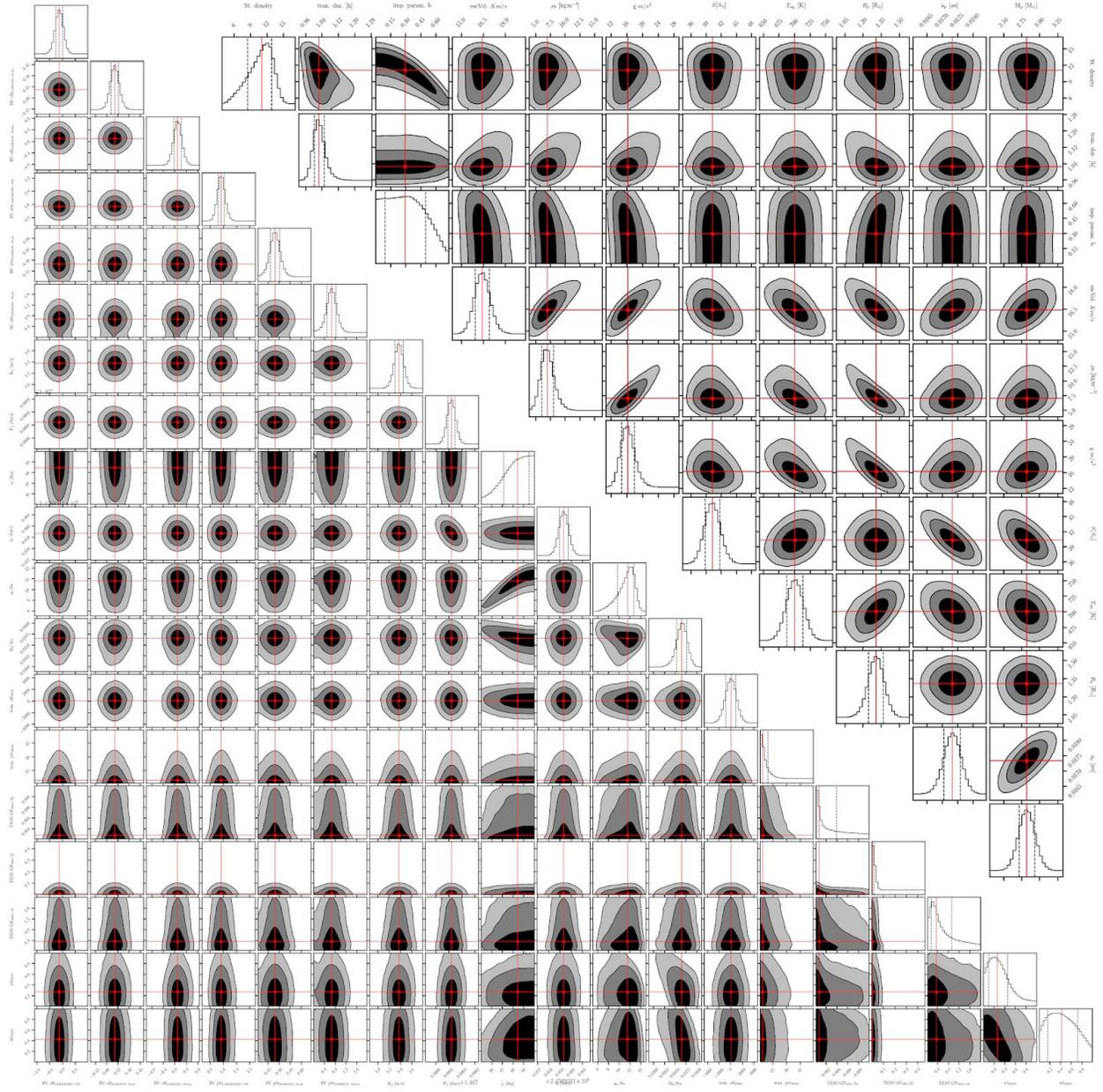

**Fig. S8. Results of the CMT model fitting.** Lower left correlation plot shows the global parameter posterior probability distributions from the nested sampling analysis. Upper right corner shows physical parameters derived from the fitted parameters. The position of the median of each posterior probability distribution is marked with red grid lines. The black contours on the 2D panels represent the 1σ, 2σ, and 3σ confidence levels of the overall posterior samples. The panels on each diagonal show the 1D histogram distribution of each parameter, while the dashed black lines show the 68.3% percentiles. Numerical results are listed in Table S6.

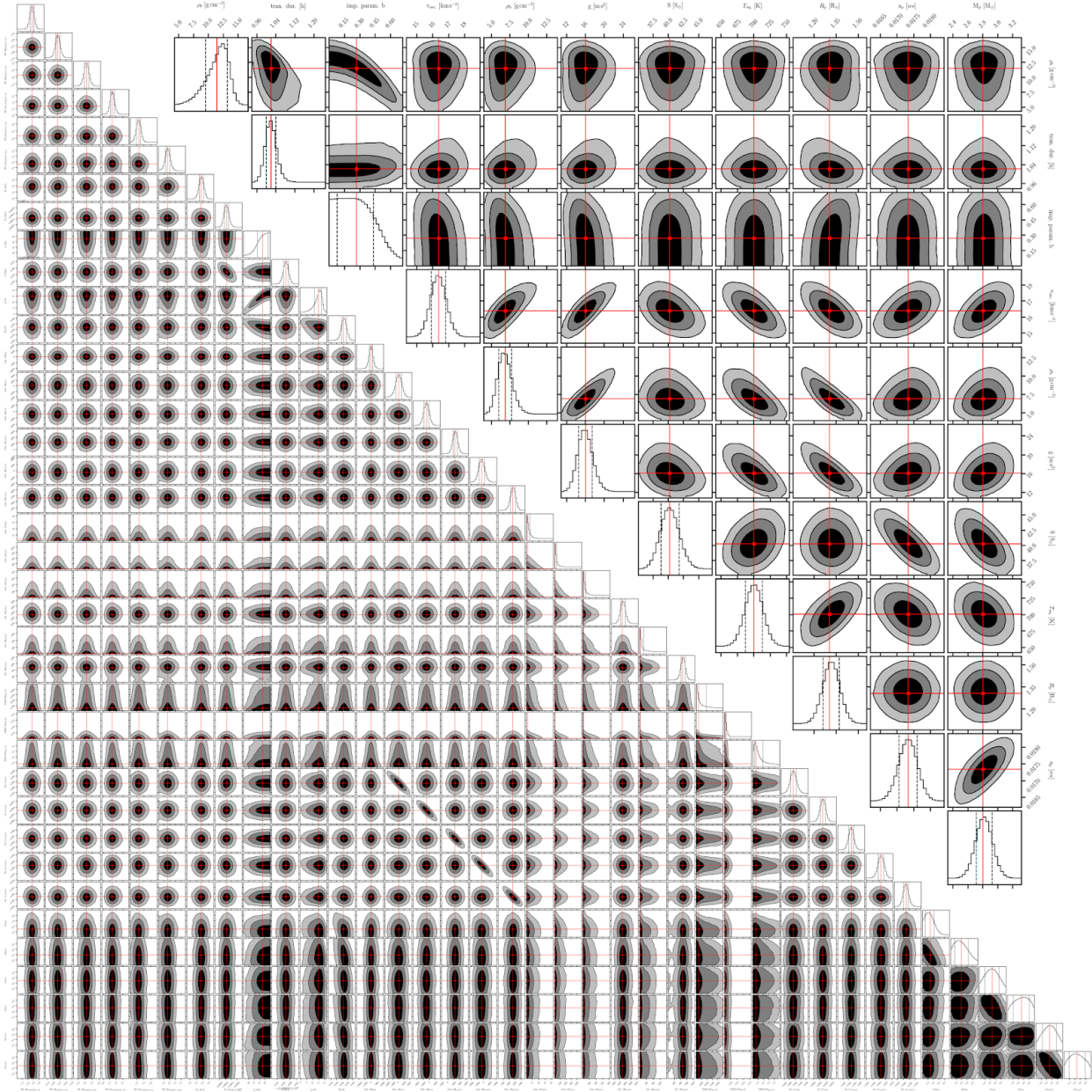

**Fig. S9. Same as Fig. S8, but for the CMT+LM model.**

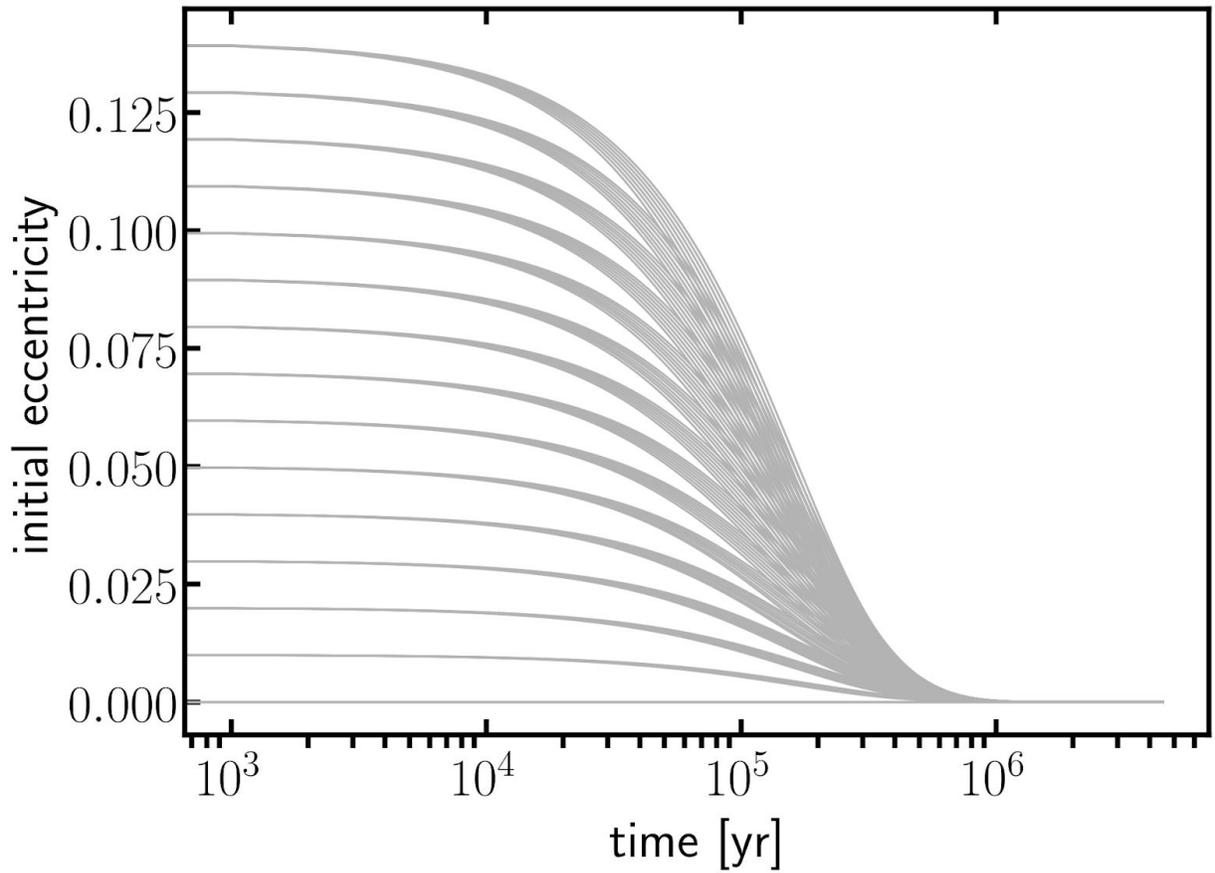

**Fig. S10. Eccentricity evolution of Gliese 486 achieved via planet-star tidal simulations.** The trajectories are constructed for various sets of initial eccentricities and semi-major axes near the best-fit of Gliese 486 b. All eccentricity trajectories converge to a circular orbit within one million year.